\documentclass[]{aa}
\usepackage{graphicx}

\begin{document}
\title{Coronal heating in multiple magnetic threads}
\author{K. V. Tam\inst{1,2} \and A. W. Hood\inst{1} \and P. K. Browning\inst{3} \and P. J. Cargill\inst{1, 4}}
\institute{School of Mathematics and Statistics, University of St. Andrews, St. Andrews, Fife, KY16 9SS, U.K.
\and 
Space Science Institute, Macau University of Science and Technology, Avenida Wai Long, Taipa, Macau 
\and
Jodrell Bank Centre for Astrophysics, School of Physics and Astronomy, University of Manchester, Manchester M13 9PL, UK
\and
Space and Atmospheric Physics, The Blackett Laboratory, Imperial College, London SW7 2BW, UK
\and
\email{awh@st-andrews.ac.uk}
}

\abstract
{Heating the solar corona to several million degrees requires the conversion of magnetic energy into thermal energy. In this paper, we investigate whether an unstable magnetic thread 
within a coronal loop can destabilise a neighbouring magnetic thread.}
{By running a series of simulations, we aim to understand under what conditions the destabilisation of a single magnetic thread can also trigger a release of energy in a nearby thread.}
{The 3D magnetohydrodynamics code, \textit{Lare3d}, is used to simulate the temporal evolution of coronal magnetic fields during a kink instability and the subsequent relaxation process. 
We assume that a coronal magnetic loop consists of non-potential magnetic threads that are initially in an equilibrium state.}
{The non-linear kink instability in one magnetic thread forms a helical current sheet and initiates magnetic reconnection. The current sheet fragments, and magnetic energy is released throughout that thread. We find 
that, under certain conditions, this event can destabilise a nearby thread, which is a necessary requirement for starting an avalanche of energy release in magnetic threads.}
{It is possible to initiate an energy release in a nearby, non-potential magnetic thread, because the energy released from one unstable magnetic thread can trigger energy release in nearby threads, provided that  the nearby structures 
are close to marginal stability.}

\maketitle

\section{Introduction}\label{sec:introductionpjc}
Energy release in the solar corona is now known to occur over a wide range of spatial and temporal scales from large eruptive flares, through compact flares, to microflares and the (hypothesised) nanoflares that may be responsible for heating the quiescent 
corona. At microflare energies and above ($> 10^{27}$ ergs), these events have been observed and they form a power law distribution $E^{- \delta}$ where $E$ is the inferred energy, and $\delta$ is in the range $1.4 - 1.8$ \citep{crosby1993,hannah2008}. 
Evidence for nanoflares is only now becoming clear with modern instrumentation: energies close to $10^{24}$ ergs are inferred \citep{warren2011,testa2013,cargill2014}. A task for theoretical models is to account for this range of event sizes, with the 
added caveat that larger events should be more \lq difficult\rq\  to trigger than smaller ones.

The largest solar flares are commonly discussed in terms of an associated eruption and ejection of plasma from the Sun (e.g. \cite{priest2000} and \cite{nishida2009}). However, the applicability of this scenario to small (compact) flares, microflares and 
smaller events is questionable, and other mechanisms need to be identified. One such fundamental process is the kink instability of a twisted magnetic field that forms a loop-like structure in the corona. This is an \lq ideal\rq\  MHD instability and 
it has fast (Alfv\'enic) growth 
during its initial phase. {There is observational evidence of the kink instability \citep{liu2007,liu2009,nandy2008,srivastava2010} along with simulations of its development in a single 
magnetic loop \citep{linton1996,galsgaard1997,baty1998,gerrard2004}.} A very attractive feature of its non-linear evolution is the rapid formation and dissipation of multiple regions of 
intense current \citep{browning2003,browning2008,hood2009,bareford2011}. Such non-linear development of magnetic reconnection can result in a turbulent state 
\citep{browning2013} that is also a good facilitator of particle acceleration \citep{gordovskyy2013}. 
The field subsequently relaxes to a lower energy state, a process that can be described by helicity-conserving, Taylor
relaxation \citep{taylor1974,browning2003,browning2008,bareford2013}. The energy released in the instability is of the order of $10^{28}$ ergs, which is characteristic of a microflare or a nanoflare storm. 
It has been shown that random footprint driving within a single loop leads to a temporal sequence of heating events of various sizes, which can
take a power law form \citep{bareford2010,bareford2011}.

It is likely that the magnetic footpoints of a large coronal loops are not located in a single magnetic source but that field lines begin and end in several separate photospheric sources. This is the idea used
in the \textit{coronal tectonics} models of \cite{priest2002}, \cite{mellor2005} and \cite{browning1986}. The tectonics model assumes that simple photospheric motions create current sheet along the separatrix surfaces
that form
between the sources. Hence, large coronal loops will have many current sheets within them and magnetic reconnection will produce heat at these locations. This suggests
that individual threads (identified by the current sheets) will be heated. A second approach is to assume that photospheric motions braid the magnetic field in a complex pattern. The magnetic field will
try to relax towards an equilibrium that may contain current sheets, as discussed by \cite{pontin2011}. {{Whether the currents in a randomly stressed magnetic field result in a smooth
equilibrium or they collapse to singular current sheets, as discussed by \cite{parker1972}, is still debated. For example, a field that should produce singular current sheets has
been shown by \cite{candelaresi2015}, using an ideal MHD relaxation method, to consist of well resolved current layers in equilibrium.}} A third approach is the individual magnetic sources could be rotated by 
photospheric vortex motions \cite{demoortel2006a,demoortel2006b}.

{All of the above approaches indicate that the coronal magnetic field is non-potential and that} an actual loop may consist of many closely packed, stressed magnetic threads. 
Now the question arises as to whether one ideally unstable thread can destabilise a nearby stable thread.  {An example of one event destabilising neighbouring magnetic fields was investigated by \cite{torok2011}.
They investigated how the onset of a highly dynamic coronal mass ejection might trigger others in the adjoining fields.} 

This paper presents an initial study using 3D MHD simulations to demonstrate how an individual unstable magnetic element can lead to the 
destabilisation of neighbouring, ideally stable elements, and so constitutes a first \lq proof of principle\rq\  that 
energy release {in a non-erupting magnetic field} can spread {across the volume of a larger structure}. Our demonstration consists of an initial state of two twisted magnetic threads (or strands or flux tubes or flux ropes), one of which is unstable to the kink instability, whereas its neighbour is stable.
{As indicated above, there are several ways of generating twisted magnetic fields: (i) flux emergence that can bring already twisted fields into the corona; (ii) relaxation of fields with non-zero relative magnetic helicity; (iii) sunspot rotation; and (iv) vortex motions at supergranular downflows. 
We assume that the net result is two twisted
threads. This assumption is made to ensure that the initial state is both non-potential and in force balance. One of these threads is then twisted further to exceed the marginal stability threshold. The amount of twist injected beyond the marginal value depends on the length of time and speed of rotation, the length of the 
thread and the growth rate of the instability.  Once all boundary motions have stopped, the subsequent evolution of the non-potential
field only arises in response to coronal disturbances.}  Following the work of \cite{hood2009}, the unstable thread is given a small initial velocity disturbance, so that the instability will grow and develop a 
complex current sheet structure. Subsequently, reconnection will occur, releasing magnetic energy and heating this thread. 

{\cite{gold1960} were the first to consider the interaction of two neighbouring magnetic loops. They argued that moving the loops together could release sufficient magnetic energy to explain a solar flare. The axial fields of the two loops were of opposite sign, in contrast with the fields modelled here, and so aided reconnection. \cite{kondrashov1999} investigated the interaction of two magnetic loops. Both loops have non-zero total axial current and so they are not in equilibrium. They are immediately attracted to
each other. In our case, we consider two magnetic threads with zero total axial current that are in equilibrium. \cite{milano1999} use Reduced MHD to investigate the formation of quasi-separatrix layers during the interaction
of two loops. \cite{mok2001} investigated how a newly emerging loop interacts with a pre-existing magnetic loop, as a model for a small solar flare. Unlike the above MHD models, in which the flux tubes 
are actively pushed together, we will investigate the interaction of two flux ropes in more detail, using full MHD simulations of magnetic fields that are initially in equiilbrium.}

The questions we wish to address include: can an unstable thread trigger {energy release in a} stable one, does the relative rotation of the magnetic field in the two threads matter, how far apart do the threads need to be to suppress any spreading of the instability and 
how close to marginal stability does the stable thread need to be for it to be triggered? How much free energy can be released as heat? We restrict attention to two threads because the nonlinear evolution of the kink instability requires high numerical resolution in order to conserve total energy and to 
resolve the current sheets created. Hence, there are insufficient computing resources at the present time to investigate, in the required detail, a truly multi-threaded loop, though this will be presented in subsequent papers.

The rest of the paper is structured in the following manner. Section 2 describes the MHD equations used and the numerical method for their solution, together with the initial equilibria and boundary conditions used. Section 3 presents the main 
results for the evolution of the energetics, the fieldline evolution, the energy dissipation and the resulting heating. Finally, a discussion of the results and our conclusions are given in the last section.

\section{Numerical method and initial equilibrium}\label{sec:method}
\subsection{MHD equations}
To understand if the relaxation of an unstable loop can result not only in heating a particular unstable thread but also in the release of magnetic energy in a neighbouring thread, we have run a variety simulations using the 3D MHD code, 
\textit{Lare3d} \citep[see][]{arber2001}. \textit{Lare3d} solves numerically the MHD equations
\begin{eqnarray}
\frac{\partial \rho}{\partial t} &=& -\nabla \cdot \left(\rho \textbf{v} \right), \label{continuity_eqn} \\
\frac{\partial}{\partial t} \left(\rho \textbf{v} \right) &=& -\nabla \cdot \left(\rho \textbf{v} \textbf{v} \right) + \frac{1}{\mu_0} \left(\nabla \times \textbf{B} \right) \times \textbf{B} - \nabla P  +\nabla \cdot {\bf S}, \label{momentum_eqn}  \\
\frac{\partial \textbf{B}}{\partial t} &=& \nabla \times \left(\textbf{v} \times \textbf{B} \right) - \nabla \times \left(\eta \frac{\nabla \times \textbf{B}}{\mu_0} \right),  \label{induction_eqn} \\
\frac{\partial}{\partial t} \left(\rho \epsilon \right) &=& -\nabla \cdot \left(\rho \epsilon \textbf{v} \right) - P \nabla \cdot \textbf{v} + \eta \textbf{j}^2 + Q_{visc}, \label{energy_eqn}
\end{eqnarray}
with internal energy density, $\epsilon$, given by
\begin{eqnarray}
\epsilon = \frac{P}{\left(\gamma - 1 \right)\rho}.
\end{eqnarray}
$\textbf{{v}}$ is the plasma velocity, $\textbf{B}$ the magnetic field, $\textit{P}$ the gas pressure, $\gamma = 5/3$ the ratio of specific heats, $\rho$ the mass density, $\mu_0$ 
the magnetic permeability and $\eta$ the resistivity. The viscous heating term is $Q_{visc} = \epsilon_{ij}S_{ij}$, where $\epsilon_{ij}$ is the strain rate and 
${\bf S} = S_{ij}$ is the viscous stress tensor, as discussed by \cite{arber2007}. The shock viscosities, used to ensure that
the Rankine-Hugoniot relations are satisfied across shocks, result in shock heating and this must be added to the viscous heating term at the correct location.

In this study, we ignore the effects of gravity, thermal conduction and optically thin radiation, which are included in \cite{botha2011} and Bareford et al (in prep). The application of this code to relaxation theory has been discussed in \citet{browning2008} and \citet{hood2009}.

The equations in \textit{Lare3d} are made dimensionless as follows
\begin{eqnarray*}
&\textbf{\textit{r}}& \rightarrow \textit{r}^{*} \tilde{\textbf{\textit{r}}}, \quad
\textbf{B} \rightarrow B^{*} \tilde{\textbf{B}}, \quad
\textbf{v} \rightarrow v_A \tilde{\textbf{v}}, \\
&\textit{P}& \rightarrow \textit{P}^{*} \tilde{\textit{P}}, \quad
\textit{t} \rightarrow \textit{t}^{*} \tilde{\textit{t}}, \quad
\rho \rightarrow \rho^{*} \tilde{\rho},
\end{eqnarray*}
where a tilde denotes a dimensionless variable. $v_A = B^{*}/\sqrt{\mu_0 \rho^{*}}$ is the typical Alfv\'en speed, $\textit{t}^{*} = \tau_{A} =  \textit{r}^{*} / v_A$ is the Alfv\'en transit time 
across the thread and $\textit{P}^{*} = B^{{*}{2}}/ \mu_0$. Here, $\textit{r}^{*} = \textit{R}$ is the thread radius, and $B^{*} = B_0$ is the 
maximum value of the initial axial field. The electric current is expressed in units of $B^{*}_0/ \mu_0 \textit{r}^{*}$ and the dimensionless temperature is determined from
the ideal gas law for a fully ionised Hydrogen plasma
\begin{eqnarray}
\tilde{\textit{P}} = 2 \tilde{\rho} \tilde{\textit{T}}.
\end{eqnarray}
In addition to the variables in Equations (\ref{continuity_eqn}) - (\ref{energy_eqn}), the magnetic, kinetic and internal energies are also calculated. 
A dimensionless resistivity, $\eta$, essentially the inverse Lundquist number, is obtained by setting $\eta^{*} = \mu_0 \textit{r}^{*} v_A$. The resistivity is not assumed uniform and we take
\begin{eqnarray}
 \eta = \eta_b + \left\{
\begin{array}{l l}
        \eta_0, & \quad |\textbf{j}| \geq j_{crit}, \\
        0, & \quad |\textbf{j}|<j_{crit}. \\
\end{array} \right.
\end{eqnarray}
where $\eta_0 = 10^{-3}$ is the anomalous resistivity and $\eta_b = 0$ is the background value. The anomalous resistivity is only switched on when the magnitude of the current exceeds a 
critical value of $j_{crit}=5$. This value is chosen to be greater than the maximum of the equilibrium current, (see below), ensuring that the anomalous resistivity is only switched on when a current
sheet is forming.

Each equilibrium thread has a normalised radius of unity. The length of each thread is taken as 20 so that the width to length ratio is 10. This value of
the inverse aspect ratio is, of course, a compromise between modelling very narrow magnetic threads and computational resolution.
The computational resolution is discussed below.

\subsection{Initial equilibria}
Consider the coronal situation, where the ratio of the gas pressure to the magnetic pressure is so small (around $10^{-3}$), then the magnetic field can be assumed to be force-free. 
Hence, $\nabla \times \textit{\textbf{B}} = \alpha \textit{\textbf{B}}$. 
For simplicity, we assume that each equilibrium magnetic thread can be modelled by a straight twisted cylinder, with the cylinder axis located at $(x_0, y_0)$, and we use the smooth $\alpha$ profile 
that only depends on radius, $r$, as described by \citet{hood2009}. Therefore, each magnetic thread has zero net axial current. {It is expected that the threads in a coronal loop do consist of
stressed magnetic fields and our model provides a simple description of non-potential magnetic fields within each thread. By considering an idealised generic configuration, we can 
focus on the underlying physical processes involved. We do not expect that increasing the geometric complexity will significantly change the underlying physics and, indeed, \cite{bareford2015} have demonstrated
that the main features of energy release through nonlinear physics are retained in more more realistic configurations.}

For a radial coordinate defined by $r^2 = (x-x_0)^2 + (y - y_0)^2$, the magnetic field components of each thread have the form
\begin{eqnarray}
 B_{\theta} &=& \left \{ \begin{array}{cc} B_0 \lambda r (1-r^2)^3,& r < 1, \\
 0, & r>1, \end{array}\right .\label{eq:b1}\\
 B_z &=& \left \{ \begin{array}{cc} B_0 \sqrt{1 -\frac{\lambda^2}{7} +\frac{\lambda^2}{7} (1-r^2)^7 -\lambda^2 r^2 (1-r^2)^6}, & r < 1, \\
 B_0 \sqrt{1 -\frac{\lambda^2}{7}}, & r > 1, \end{array} \right .  \label{eq:b2}\\
 \alpha &=& \left \{ \begin{array}{cc} \frac{2 \lambda (1-r^2)^2 (1-4 r^2)}{B_z}, & r < 1, \\
 0, & r > 1. \end{array} \right . \nonumber
\end{eqnarray}
$B_0$ is the magnetic field strength on $r=0$ and $\lambda$ is a constant parameter related to the twist in the magnetic field. The maximum value of $\lambda$ is restricted by the fact that 
$B_z^2$ must be positive and, therefore, $\lambda < 64/965 \sqrt{1351} = 2.438$. $\lambda$ controls the stability properties of the thread and the marginal stability value, $\lambda_{crit}$
does depend on the length, $2 L_{z}$, of the thread. For our case, $2 L_{z} =20$ and $\lambda_{crit} = 1.586$.  The stability threshold for longer threads will given by smaller values of $\lambda$ \citep{bareford2010}.
$\lambda$ also controls the maximum value of the magnitude of the current, i.e. $2 \lambda B_0$. Note that $\lambda$ is positive in each thread so that they both
have the same sense of rotation.
The critical current used in switching on the anomalous resistivity is always chosen to be larger than this value. The axial flux within a thread is 
\begin{displaymath}
\Phi = 2 \pi \int_{r=0}^{r=1} B_z r dr = 2\pi B_0 f(\lambda),
\end{displaymath}
where $f(\lambda)$ is a monotonically decreasing function of $\lambda$. Thus, $B_0$ is proportional to the axial magnetic flux in the thread. Note that the values of $\lambda$ and $B_0$ are related 
for the different threads. From Equation (\ref{eq:b2}), a smaller value of $\lambda$
requires a smaller value of $B_0$ so that all threads are embedded in the same uniform potential field.

Since the $\alpha$ profile has both positive and negative values, the total magnetic helicity in the equilibrium field will be relatively small. In this case, the Taylor relaxed state will be very close to 
a potential field. Hence, in the mid-plane $z=0$, the magnetic fields will evolve to a nearly uniform field in the axial direction. $B_x$ and $B_y$ will only be significantly different from zero near the photospheric boundaries, $z=\pm L_z$.

\subsection{Boundary conditions}
For each of the numerical experiments investigated below, we have a computational domain with $2 L_x=8$, $2 L_y=4$ and $2 L_z=20$. We run the simulation on a numerical grid of
640 (in $x$) by $320^2$ (in $y$ and $z$). This maintains the same resolution in $x$ and $y$. Convergence studies were undertaken with coarser grids.
We assume that the side boundaries are given by
a perfectly conducting wall. Hence, the velocity components vanish at $x = \pm L_x$ and at $y = \pm L_y$. Obviously the Sun does not have perfectly conducting side boundaries but these
conditions result in a more stable magnetic field. Hence, any instability that shows up in our simulations will definitely occur when alternative (and less stabilising) 
boundary conditions are used. Nonetheless, the values of $L_x$ and $L_y$ are chosen large enough to reduce the
influence of the side boundaries on the subsequent evolution. It has been shown by \cite{browning2008} that
the outer boundary does not significantly influence the stability properties of zero net current magnetic fields, if the boundary is located at a distance that is more than twice the radius of the magnetic field.

In order to simulate the anchoring of the magnetic footpoints in the dense chromosphere/photosphere \citep[see][]{hood1986}, we assume line-tying boundary conditions, namely that the velocity components vanish
at the two ends $z = \pm L_{z}$. The axial length, $2 L_z=20$, is taken as large as is computationally viable, while retaining numerical resolution. Increasing $L_z$ reduces the stablising effect of the line-tying and, again, any instability found in these simulations will occur in a longer thread. The magnetic field vector components, density and temperature are assumed free-floating at the footpoints so that\begin{eqnarray}
 \frac{\partial B_x}{\partial z} = \frac{\partial B_y}{\partial z} = \frac{\partial B_z}{\partial z}  =\frac{\partial \rho}{\partial z} = \frac{\partial T}{\partial z} = 0, \quad \textrm{at } z=\pm L_{z}.
\end{eqnarray}

\section{Numerical results}\label{sec:results}
The aim of our studies is to investigate whether an unstable thread can trigger an energy release in nearby threads, as the initial stages of an avalanche. An unstable equilibrium thread will 
always be located at right-hand side of the system. The twist parameter for the unstable thread is taken as $\lambda=1.8$,
a value beyond the ideal MHD marginal stability threshold. This thread is given an initial helical velocity perturbation \citep[see][]{hood2009} and the perturbation is chosen to be close to the form of the most unstable mode. The
instability develops and, by $t = 70 \tau_{A}$, where $\tau_{A}$ is the Alfv\'en crossing time, it is dynamically important.

{We have selected $\lambda = 1.8$ simply because it is beyond the marginal value and generates a fast enough instability for interactions with the other thread to develop within a computationally reasonable time. Previous work
\citep{gerrard2002,gerrard2004,rappazzo2013} has shown that an initially potential loop can be twisted by photospheric motions until the kink instability occurs. We can increase $\lambda$
beyond the marginal value by imposing a dimensionless rotational velocity, $V_0$, at both ends, that increases the value of $\lambda$ by approximately $70 V_0$ or, equivalently, twice the distance each footpoint rotates through in the given time 70. 
This time is a lower limit based on the time it takes the instability to develop nonlinearly. For $\lambda$ is closer to the marginal value and the growth of the instability will be slower. 
Thus, we can estimate the rotational velocity needed to reach $\lambda = 1.8$ as $0.15/70 = 0.002$ or $0.2\%$ of the background Alfv\'en speed.}

Before looking for evidence of triggering an instability in a stable thread, we must discuss how the individual magnetic threads behave when there is no initial perturbation imposed. The initial magnetic fields are analytically in
equilibrium but, because of truncation errors introduced by the finite difference approach, there will be a small but non-zero Lorentz force. The size of this force depends on
the grid resolution. Hence, due to these small truncation errors, an unstable magnetic field will eventually excite the kink instability after just over 200 $\tau_{A}$ on our high resolution grid. 
A thread is destabilised by an unstable neighbouring thread if it releases magnetic energy the time interval, $70\tau_{A} < t < 200\tau_{A}$. 

There is no such issue with the stable threads having, say, $\lambda = 1.4$.
There is no instability, due to truncation errors, during the runtime of the simulations. In such a case, any release of magnetic energy is entirely due to
the destabilisation created by the unstable neighbour. It is worth pointing out that, when $\lambda$ is close to the critical value, $\lambda_{crit}$, for marginal stability, a non-zero value for the background
resistivity, $\eta_{b} > 0$, can result in a slight diffusion of the field and current. This can generate a non-zero Lorentz force that creates a small flow and changes the radial profile of the current. Hence, to
avoid this, we assume $\eta_{b} = 0$.

From now on, we will only present results for the simulations in which the right-hand thread is disturbed by a helical velocity perturbation. This thread is placed with its axis at either $(x,y) = (2,0)$
or $(x,y) = (0,0)$.
\begin{figure}[ht]
\centering
{\includegraphics[width=0.49\textwidth]{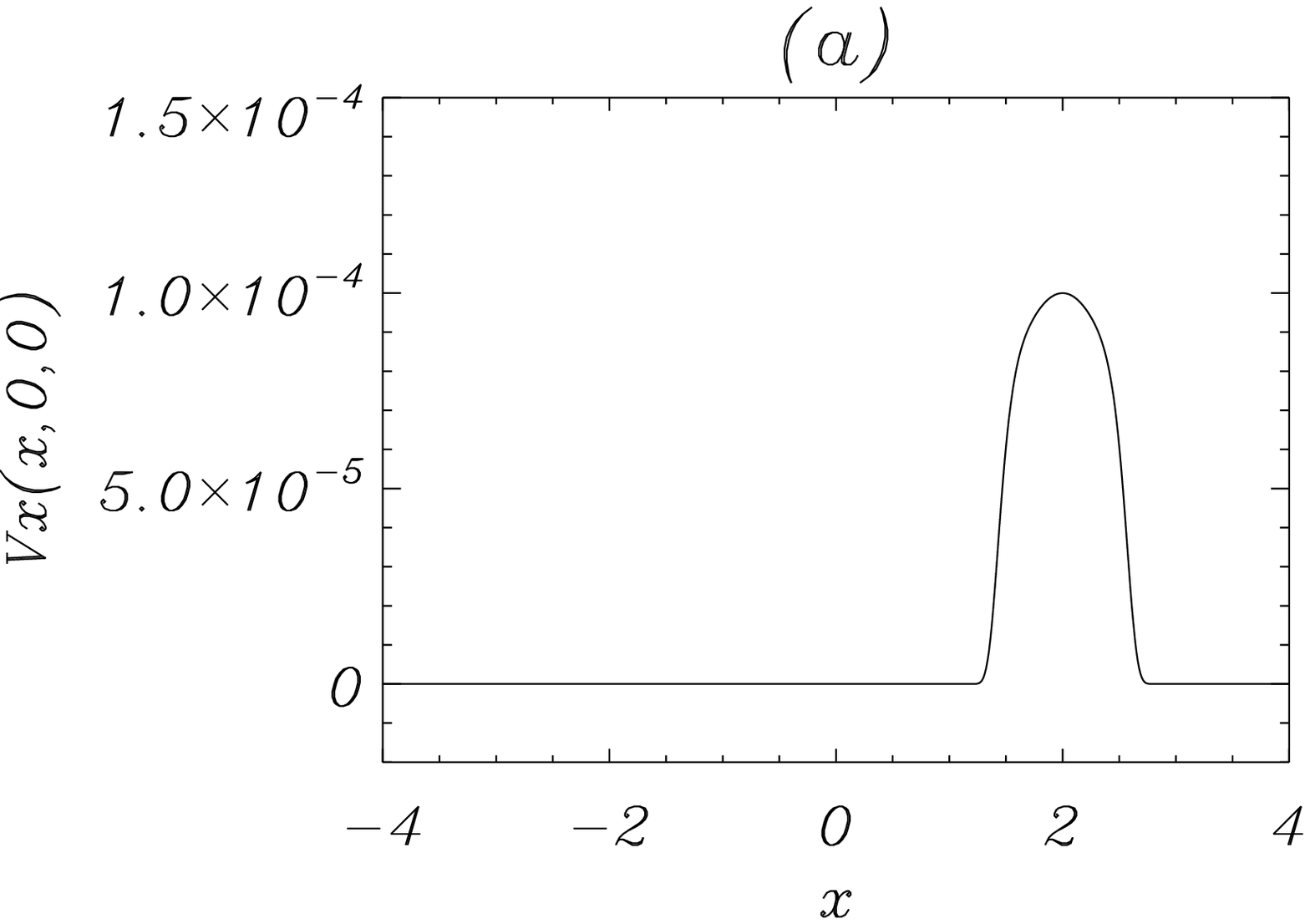}}
{\includegraphics[width=0.49\textwidth]{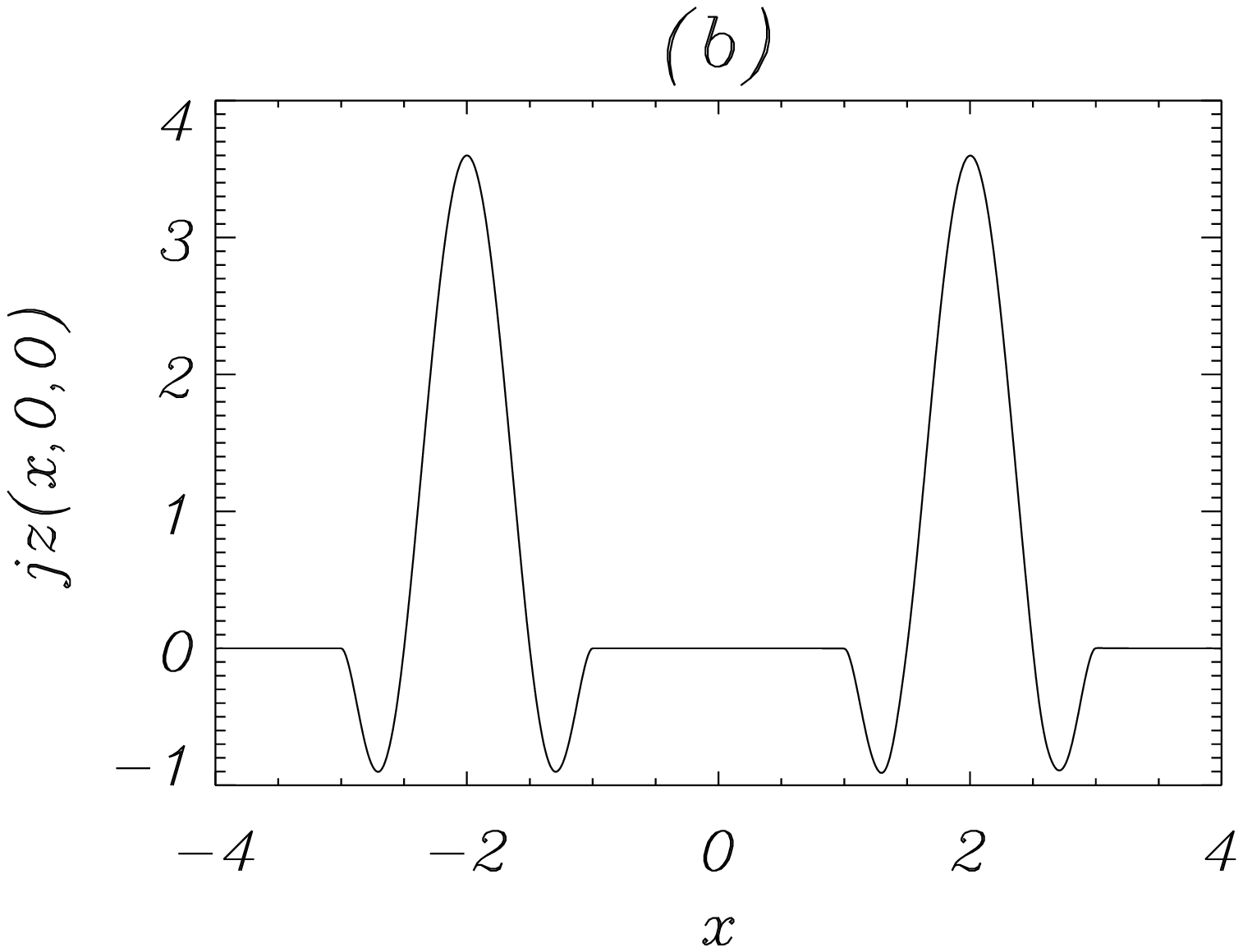}}
\caption{Initial profiles of \textbf{(a)} velocity and \textbf{(b)} axial current, $j_{z}$, at the mid-plane for Case 1.}
\label{case1_vx_j}
\end{figure}
Since this thread is always unstable, the initial perturbation will eventually grow in amplitude, allowing the kink instability to develop reasonably quickly. 
A nearby twisted thread is always placed at the left-hand side of the system, with its axis at $(x,y)=(-2,0)$. Our aim is to investigate whether the unstable thread on the right can destabilise the one on the left
or not.

We consider four different configurations of the two neighbouring magnetic threads. {The properties, parameters and results are summarised in Table~\ref{two_summaries}. As well as the locations of the threads' axes, the $\lambda$ values and whether the thread is
stablue or unstable, Table~\ref{two_summaries} presents the time at which magnetic energy starts to decrease in each case
and the changes in magnetic and internal energies by the end of simulation. The reduction in magnetic energy, as a percentage of the initial value, is also given.}
\begin{table}[ht]
\caption{Parameters and properties of the four cases considered.}
\centering
\begin{tabular}{|ccccc|}
\hline
& Case 1 & Case 2 & Case 3 & Case 4 \\
\hline
 & \multicolumn{4}{c|}{Location of centres} \\
\hline
1st thread & (2,0) & (0,0) & (2,0)& (0,0) \\
2nd thread & (-2,0) & (-2,0) & (-2,0) & (-2,0) \\
\hline
 & \multicolumn{4}{c|}{$\lambda$ values} \\
\hline
1st thread & 1.8 & 1.8 & 1.8 & 1.8 \\
2nd thread & 1.8 & 1.8 & 1.4 &1.4 \\
\hline
 & \multicolumn{4}{c|}{Stability} \\
\hline
1st thread & Unstable & Unstable & Unstable & Unstable \\
2nd thread & Unstable & Unstable & Stable & Stable \\
\hline
 & \multicolumn{4}{c|}{Energy Release Time ($\tau_A$)} \\
\hline
1st thread & 68 & 68 & 68 & 68 \\
2nd thread & 164 & 135 & - & 138 \\
\hline
& \multicolumn{4}{c|}{Changes in Energy} \\
\hline
Magnetic & -3.031 & -3.069 & -1.495 & -2.317 \\
& (1.74\%)& (1.75\%) & (0.8\%) & (1.3\%)\\
Internal & +2.813 & +2.864 & +1.401 & +2.154 \\
\hline
\end{tabular}
\label{two_summaries}
\end{table}

\begin{figure*}
\centering
{\includegraphics[width=0.4\textwidth]{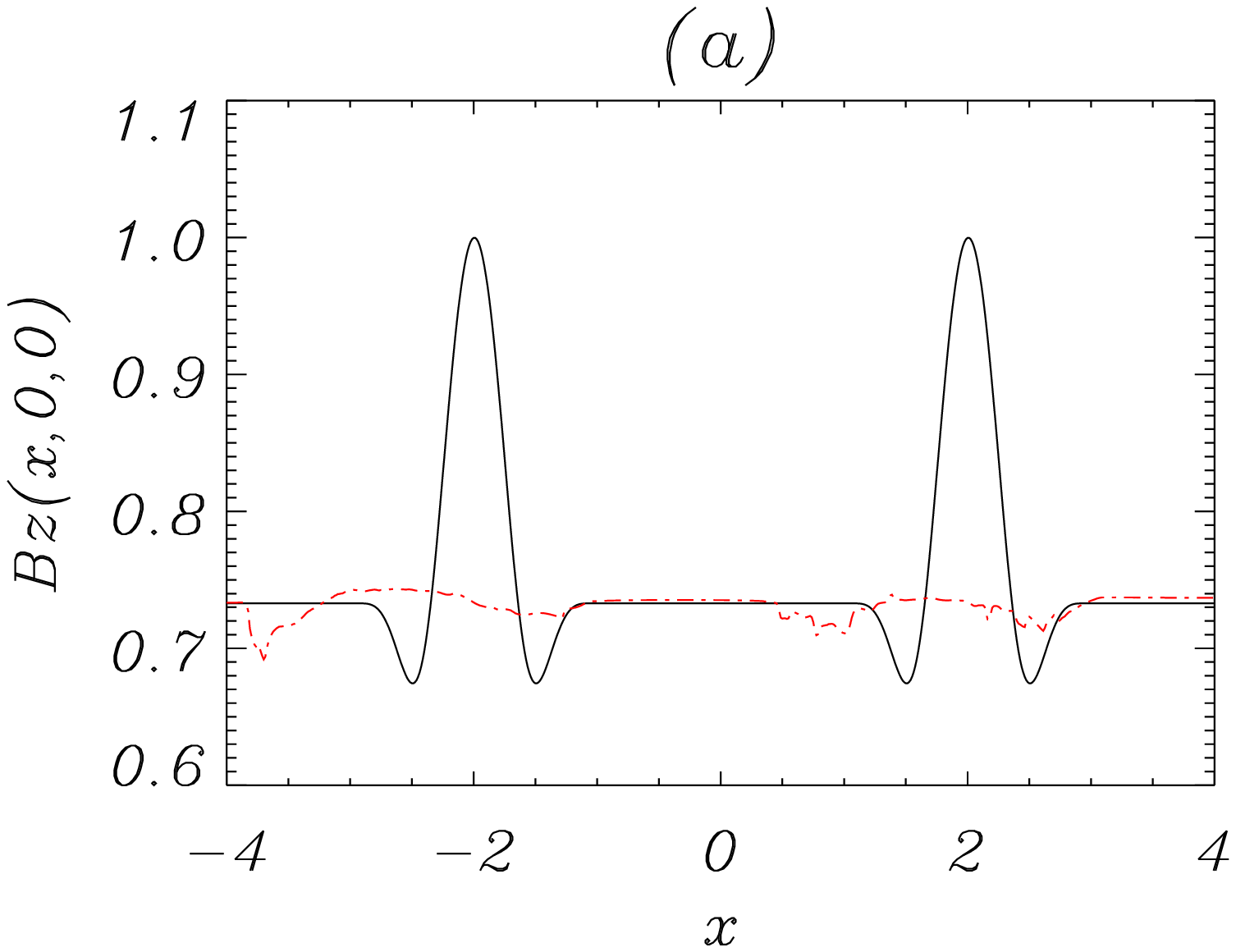}}
{\includegraphics[width=0.4\textwidth]{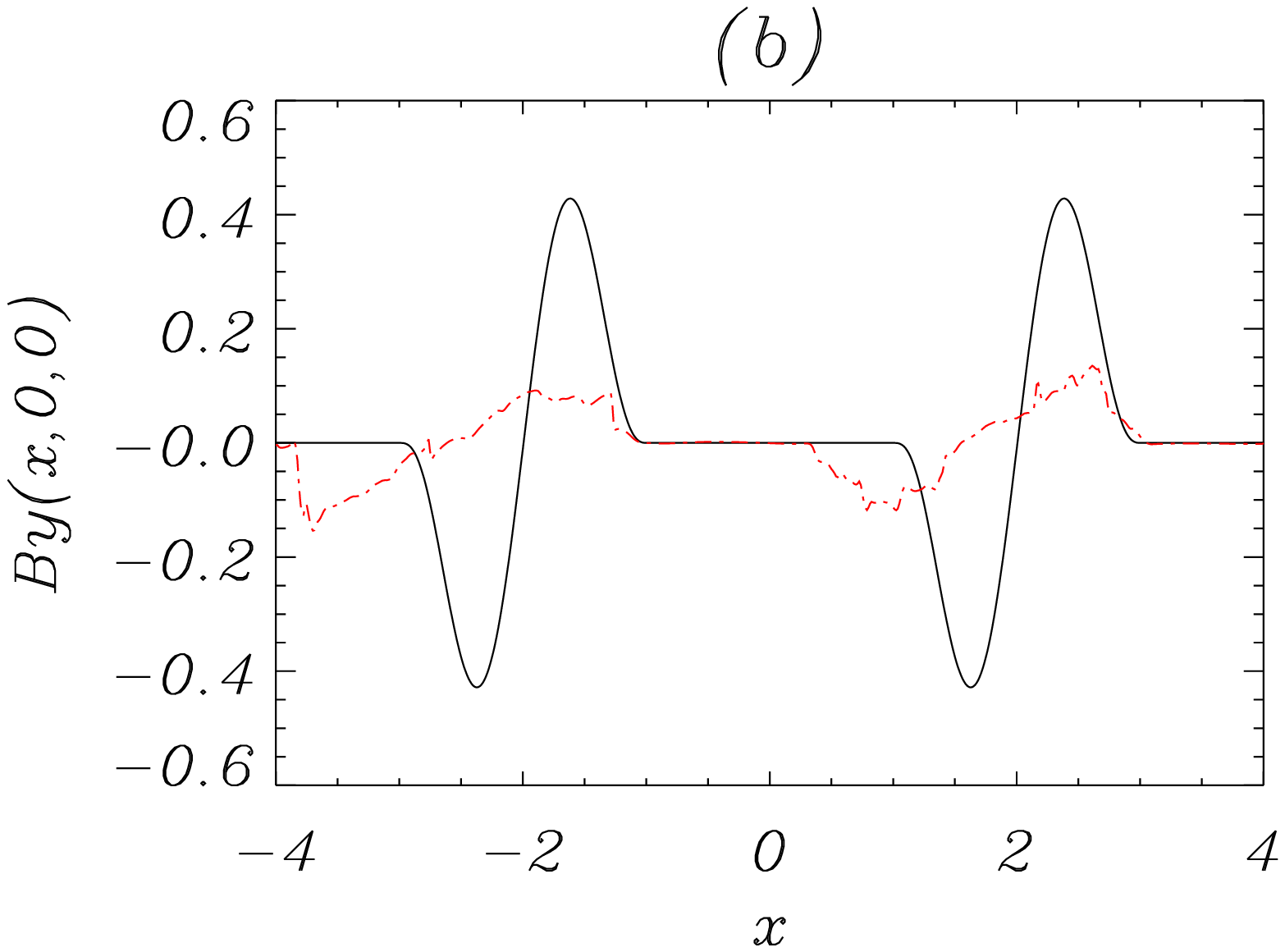}}
{\includegraphics[width=0.4\textwidth]{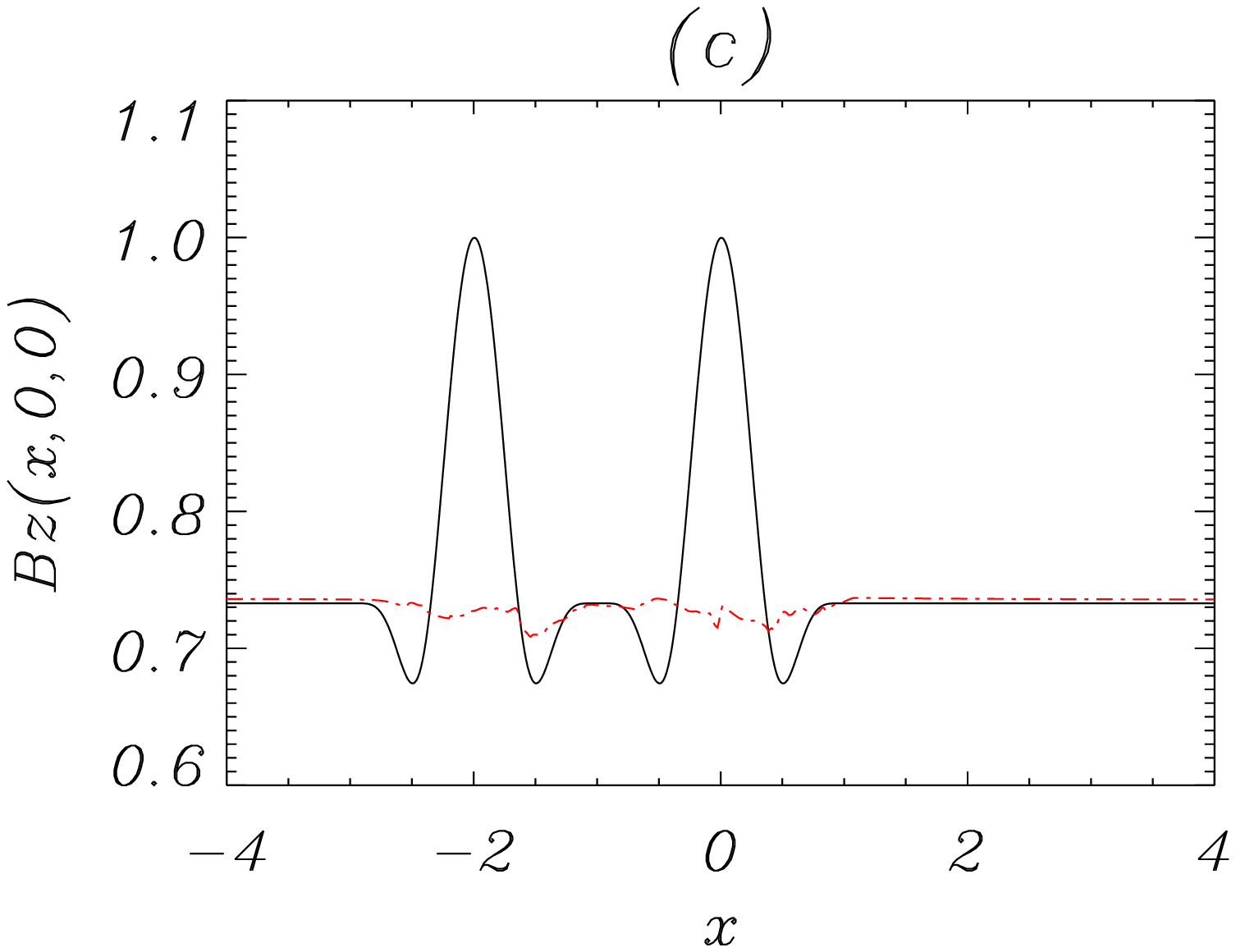}}
{\includegraphics[width=0.4\textwidth]{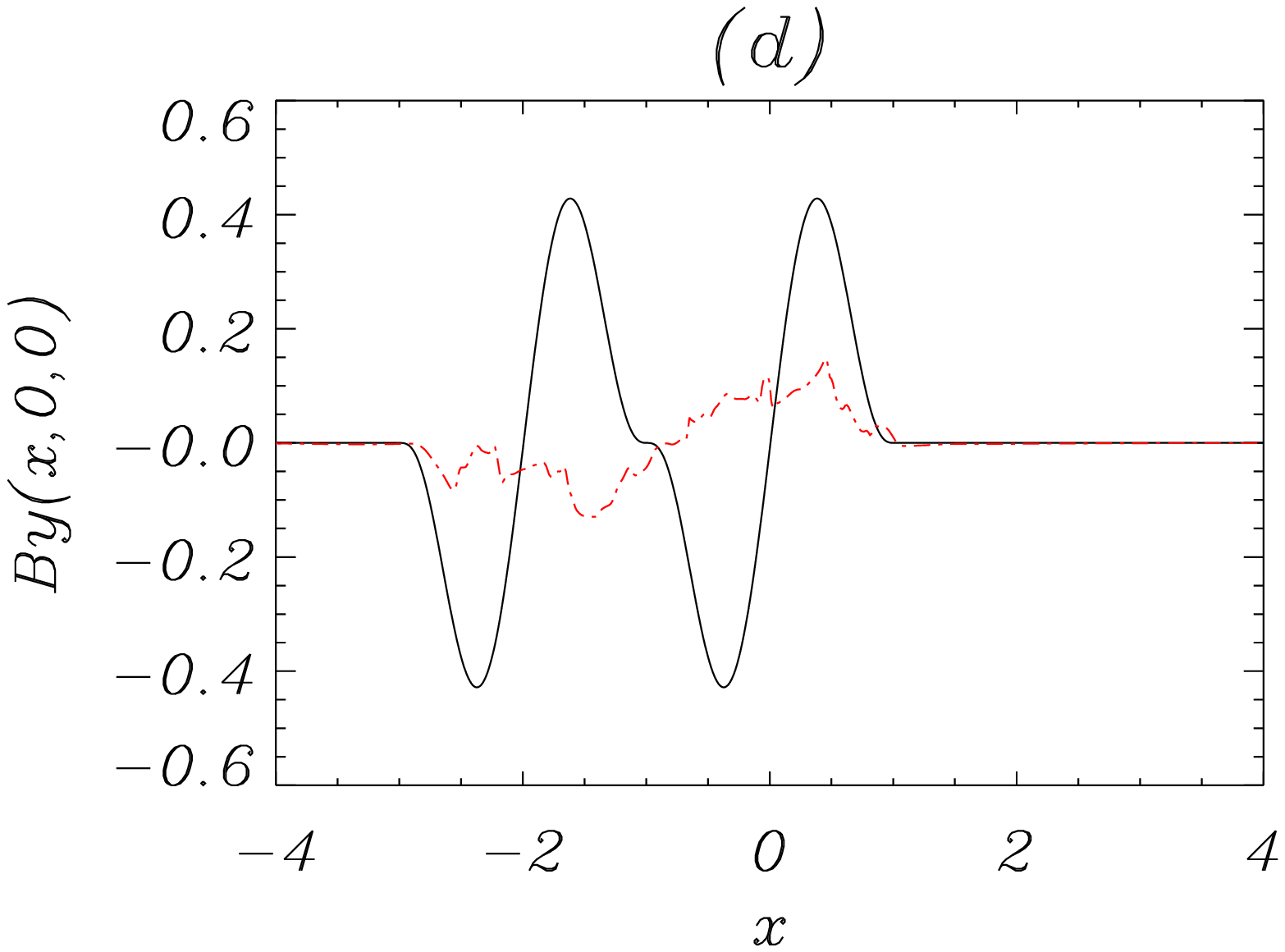}}
{\includegraphics[width=0.4\textwidth]{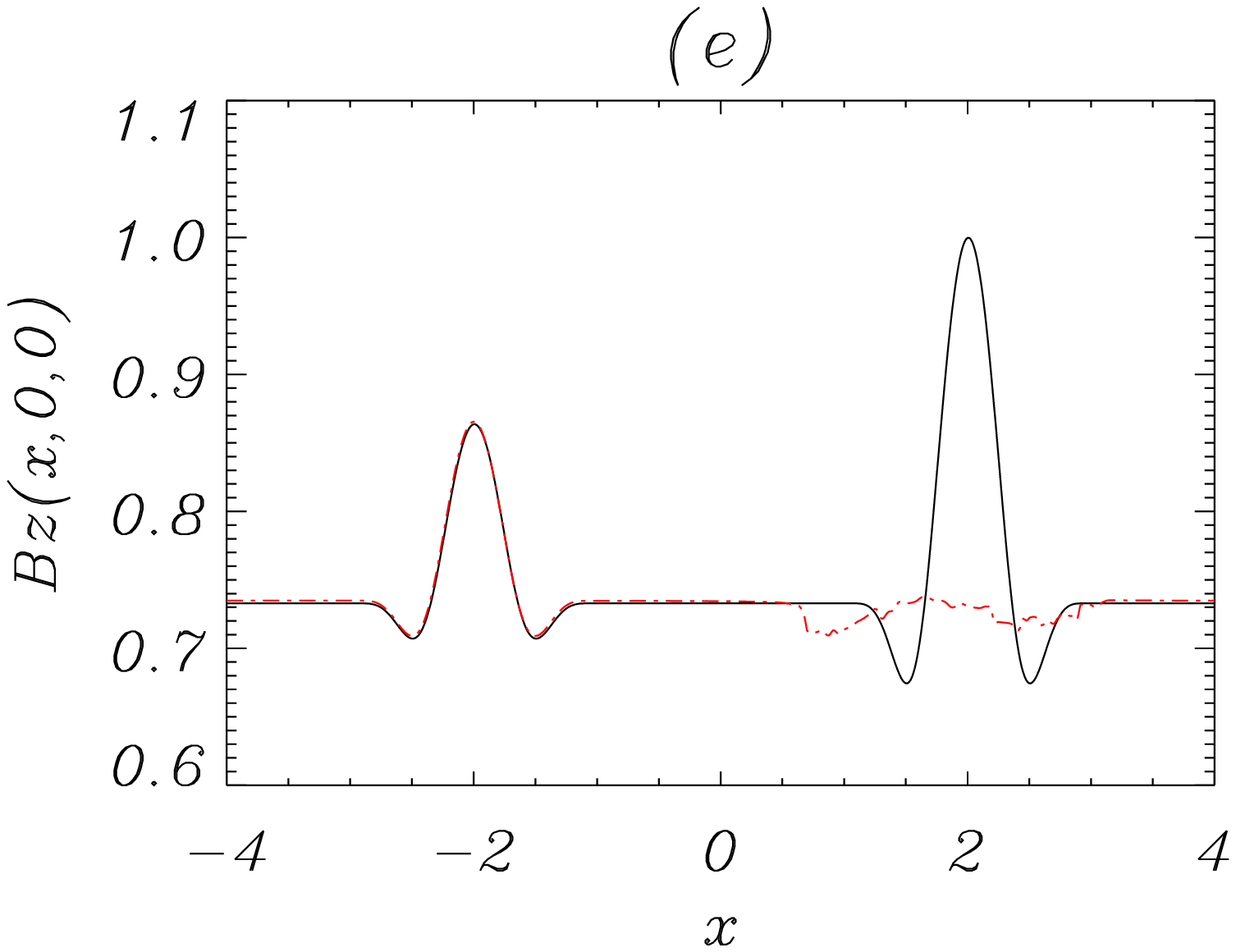}}
{\includegraphics[width=0.4\textwidth]{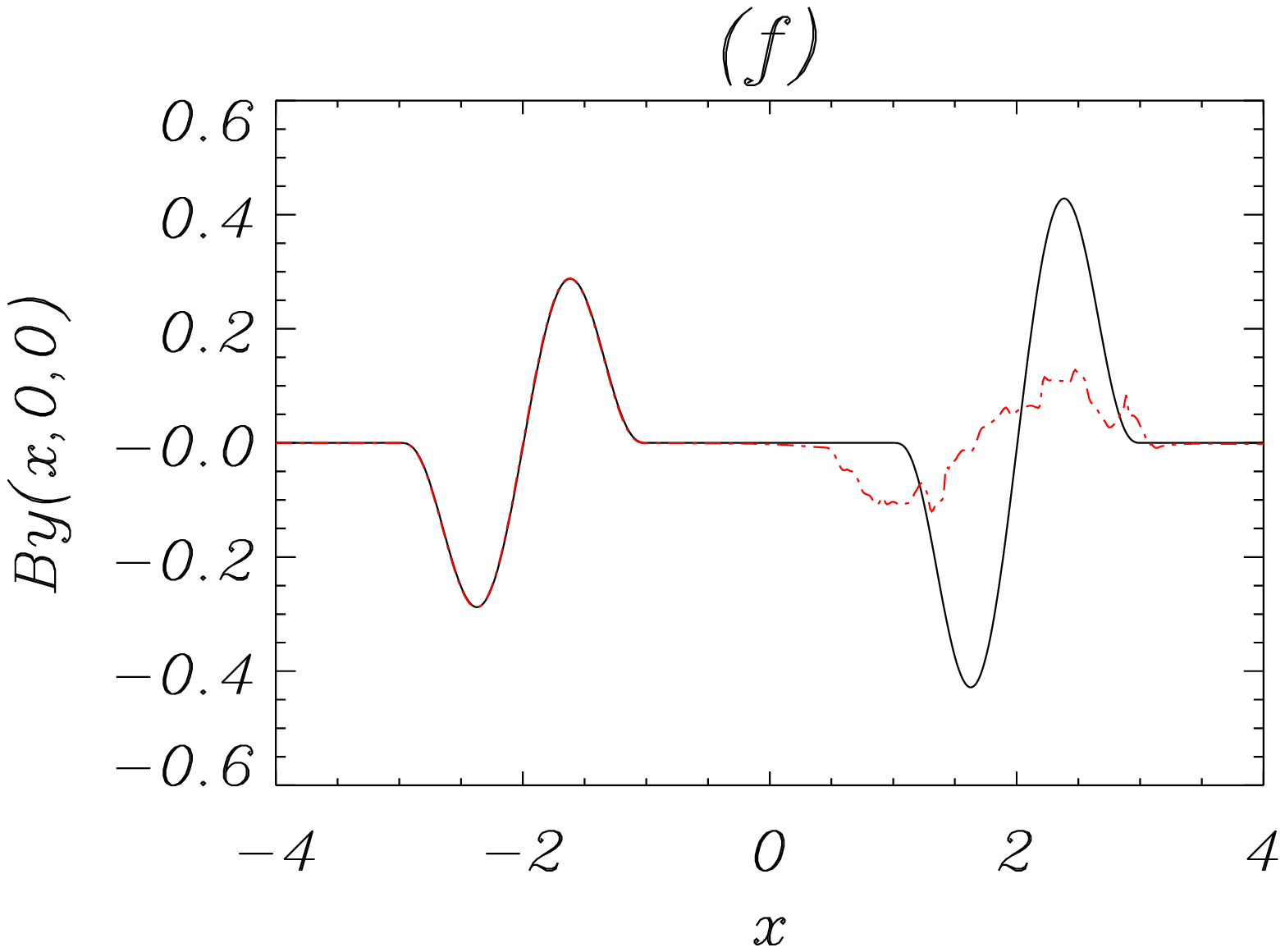}}
{\includegraphics[width=0.4\textwidth]{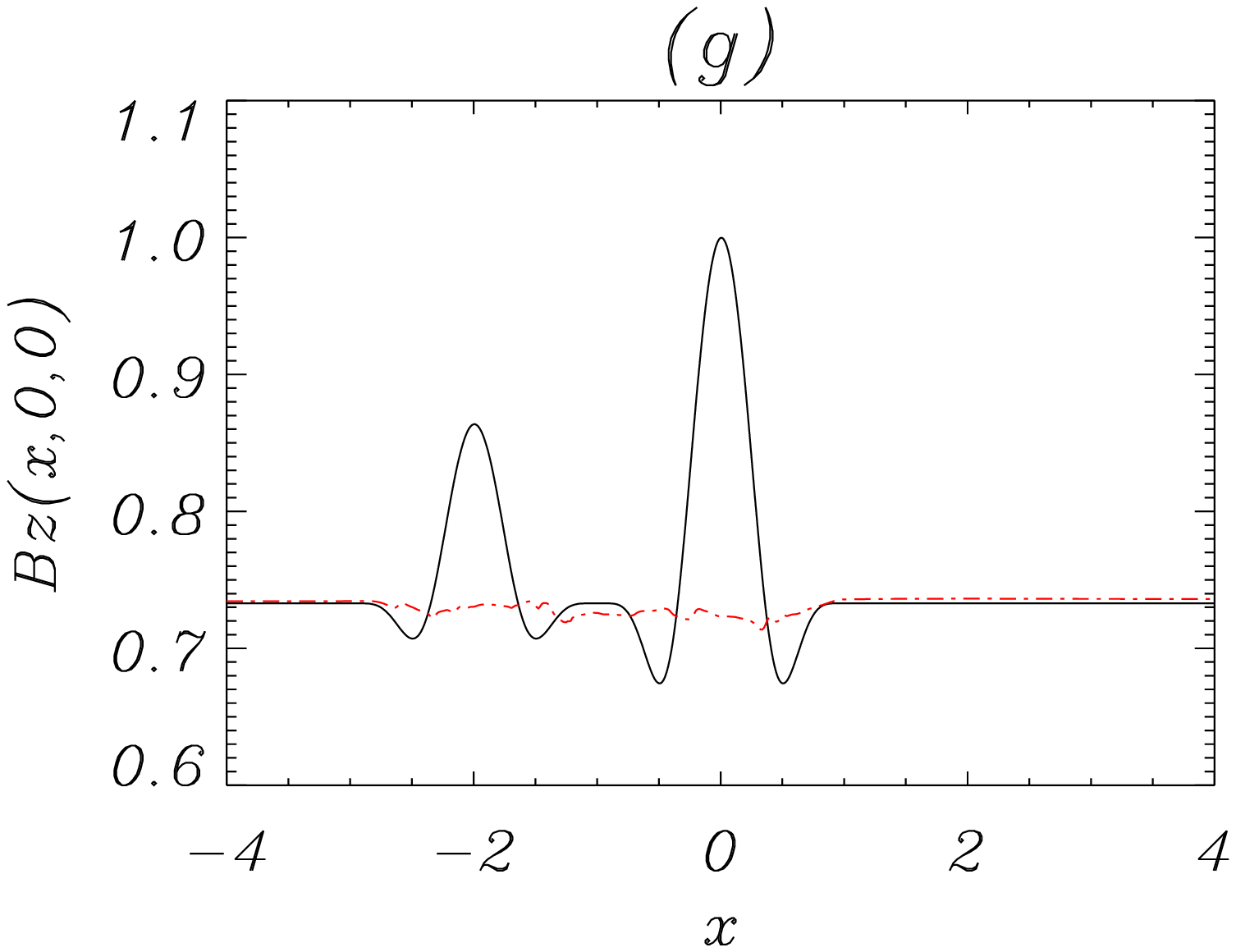}}
{\includegraphics[width=0.4\textwidth]{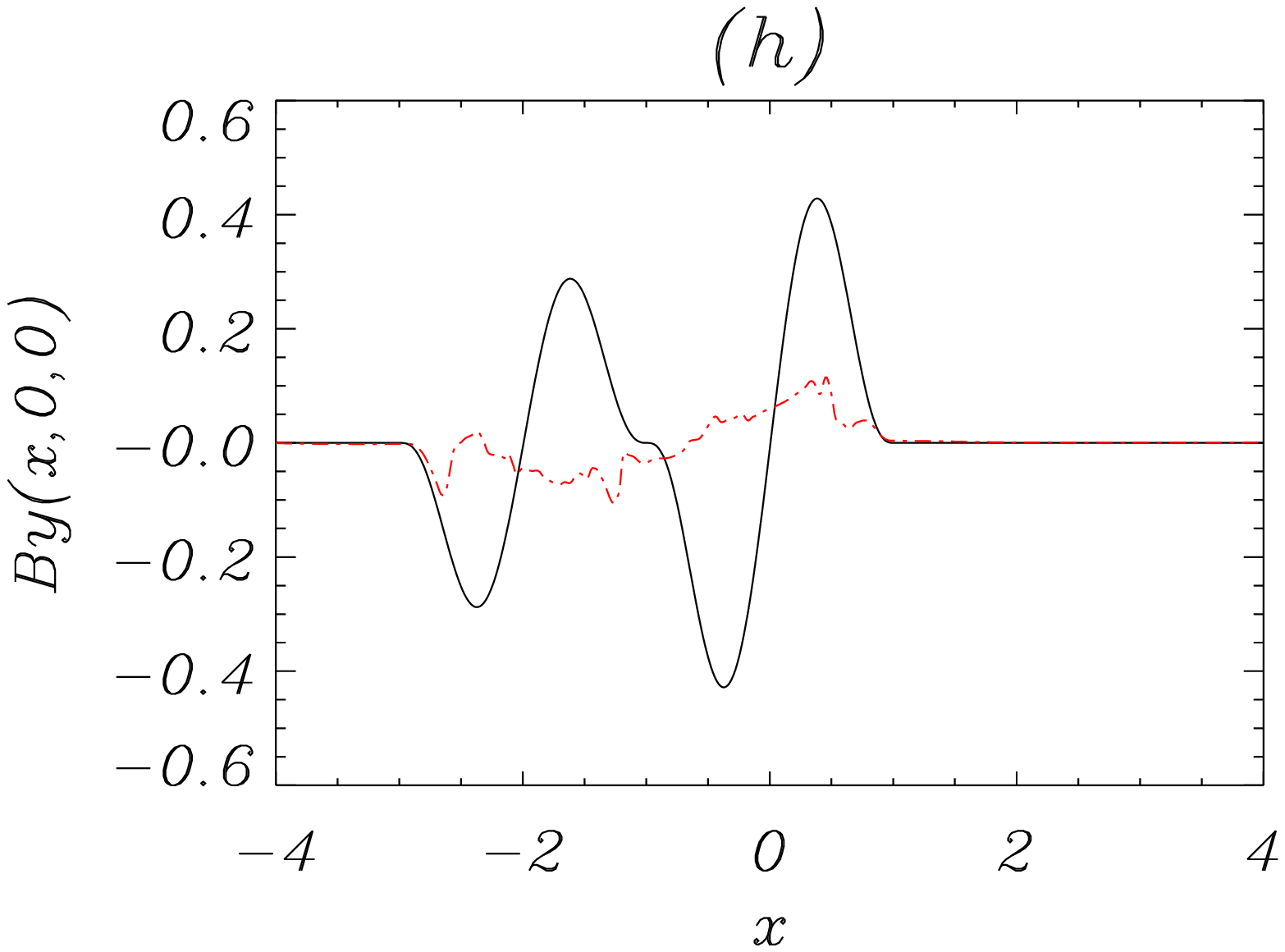}}
\caption{Initial (\textit{black}) and final at $t= 300 \tau_{A}$ (\textit{red dot-dashed}) profiles of $B_z(x,0,0)$ (left column) and $B_y(x,0,0)$ (right column) for Case 1 (top row), Case 2 (second row), Case 3 (third row) and Case 4 (bottom row).}
\label{case1_bybz}
\end{figure*}

\underline{Case 1:}
Two unstable equilibrium threads are located at $(x,y)=(2,0)$ and $(x,y)=(-2,0)$. Since the threads both have a radius of $r=1$, the outer edges of the threads are a distance 2 units from each other. 
Both fields have a twist parameter of $\lambda=1.8$. The axial current of the two threads is shown as a function of $x$ in Figure~\ref{case1_vx_j}. Only the right-hand thread is given an initial helical perturbation to the velocity. 
At the mid-plane $z=0$, this perturbation is in the $x$ direction and is shown in Figure~\ref{case1_vx_j}. Note how the perturbation
is centred on the axis of the right-hand thread and is zero outside this thread.

\begin{figure*}
\centering
{\includegraphics[width=0.98\textwidth]{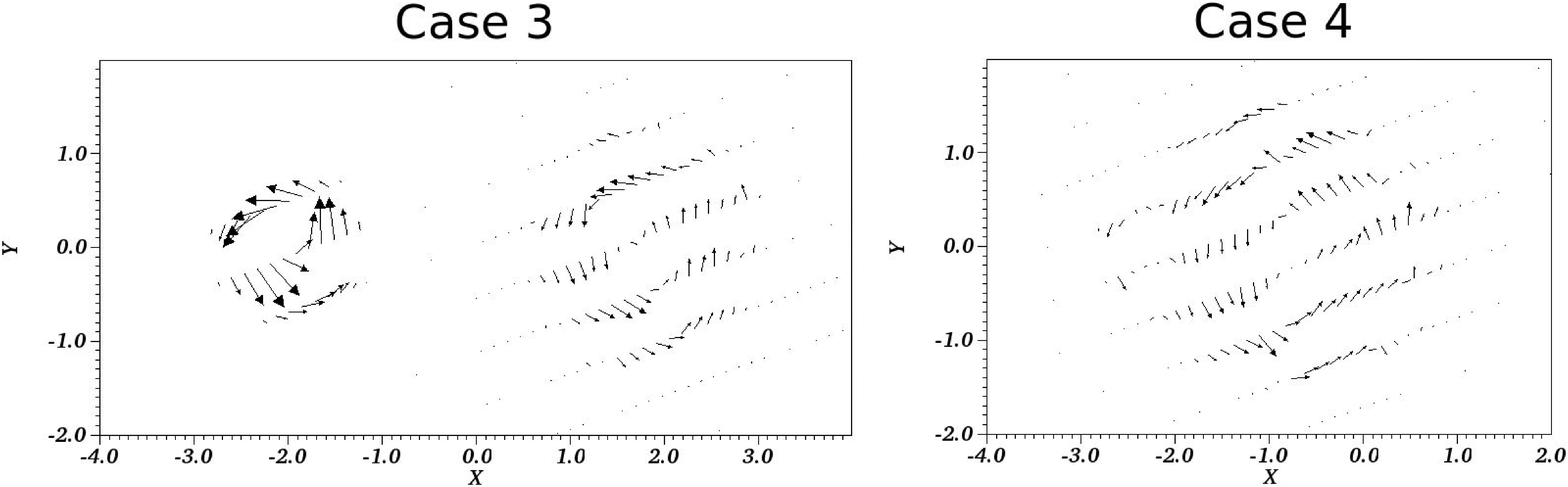}}
\caption{Projected magnetic field vectors, shown by arrows, in the midplane ($z=0$) for Cases 3 and 4 at $t=300$. The size of arrows are scaled by
factors 4 and 5 for Cases 3 and 4 respectively.}
\label{case34barrows}
\end{figure*}

\underline{Case 2:}
This is the same as Case 1, except that the threads are centred at $(x,y)=(0,0)$ and $(x,y)=(-2,0)$. Hence, the outer edges of the threads touch each other at $(-1,0)$.

\underline{Case 3:}
The axes of the two threads are at the same locations as in Case 1 but the left-hand thread has a reduced twist parameter of $\lambda=1.4$. This value of the twist parameter means that this
thread would be stable if on its own \citep{bareford2011}.

\underline{Case 4:}
This is the same as Case 3 but with the threads centred at $(x,y)=(0,0)$ and $(x,y)=(-2,0)$.

We select $B_0 = 1.0$ (see Equations~\ref{eq:b1} and \ref{eq:b2}) as the axial field strength on the axis of the unstable thread. The value of $\lambda$ of this thread is taken as $\lambda = \lambda_0=1.8$, to ensure
it is kink unstable. As mentioned above, since that the background magnetic field (at $r>1$) must be the same everywhere, 
the field strength for the stable threads with $\lambda=1.4$ and so the axial field strength in these threads is given by $B_0 = \sqrt{(1 -1.8^2/7)/(1 -1.4^2/7)} = 0.864$.

\subsection{Initial and final states}
Figure~\ref{case1_bybz} shows the initial ($t=0$) and final ($t=300 \tau_{A}$) horizontal profiles of $B_z(x,0,0)$ and $B_y(x,0,0)$ as functions of $x$ for all four cases. 
For Case 1, both threads are unstable and evolve to essentially the
same final state, with an almost uniform axial field component, $B_z$ and a significantly reduced twist in the field lines, as shown by the reduced values of $B_y$. Note how the
final diameter of each unstable thread is about 1.5 times the original value. However, the actual
temporal evolution is needed to determine whether the left-hand thread is actually driven unstable by the kink instability in the right-hand thread or occurs due to numerical truncation errors. Since
the energy release occurs before a time of $200\tau_{A}$, we can state that the left hand thread is driven by disturbances generated by the right hand one.
Note that $B_y(x,0,0)$ is clearly
zero between the two threads. Case 2 is similar, with both threads relaxing towards a more potential state. In this case, however, it appears from the $B_y$ profile that the two 
threads have combined into a single larger thread. $B_y$ is positive for $x > -1$ and negative for $x < -1$. The twist is about a common axis.

Case 3 shows just the unstable right-hand thread has relaxed, while the stable left-hand thread remains unchanged from its initial state. However, Case 4 is different. 
Both the $B_z$ and $B_y$ profiles of the left-hand thread show clear evidence of both relaxation and combining to one single magnetic structure at the end of the simulation. A stable thread is \textit{destabilised by an unstable one}.
This is our first building block for the developing an avalanche model in the MHD framework. 

Summarising, when the two magnetic threads are sufficiently close together, the instability in the right-hand thread triggers the relaxation of the left-hand thread. In addition, the two threads of Case 2 and Case 4 
have evolved into a single-loop like system by the end of the simulations. This is indicated clearly from the magnetic field vectors projected
onto the mid-plane ($z=0$) (Figure~\ref{case34barrows}). Case 3 shows the left-hand thread remains the same as its initial state while the right-hand thread has spread out over a larger area. 
Case 4 shows that the two threads have combined with the axis of the new loop located near $x=-1.0$ and $y=-0.2$, almost halfway between the two initial axes.

\subsection{The time evolution of energy}
\begin{figure*}
\centering
{\includegraphics[width=0.49\textwidth]{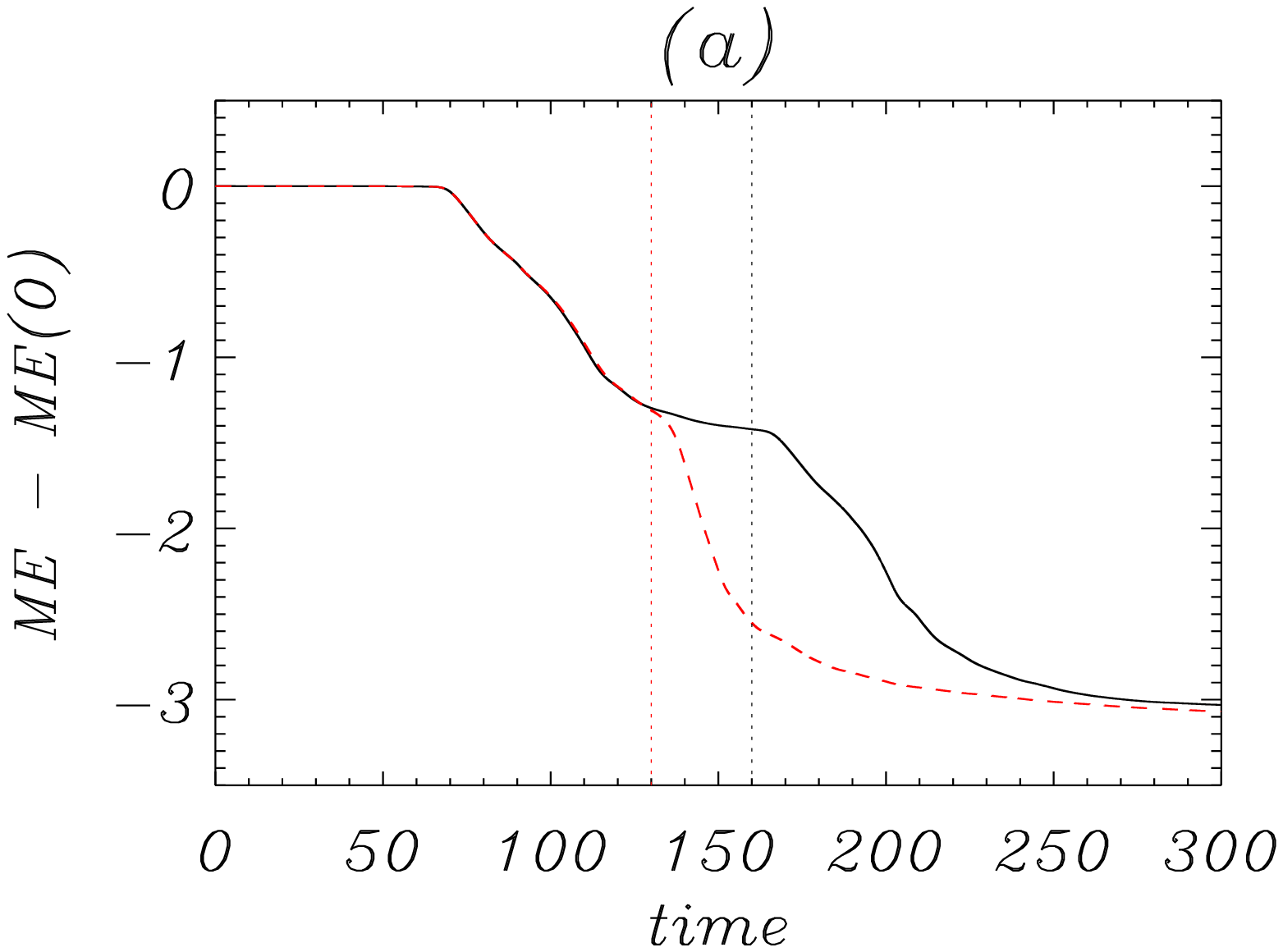}}
{\includegraphics[width=0.49\textwidth]{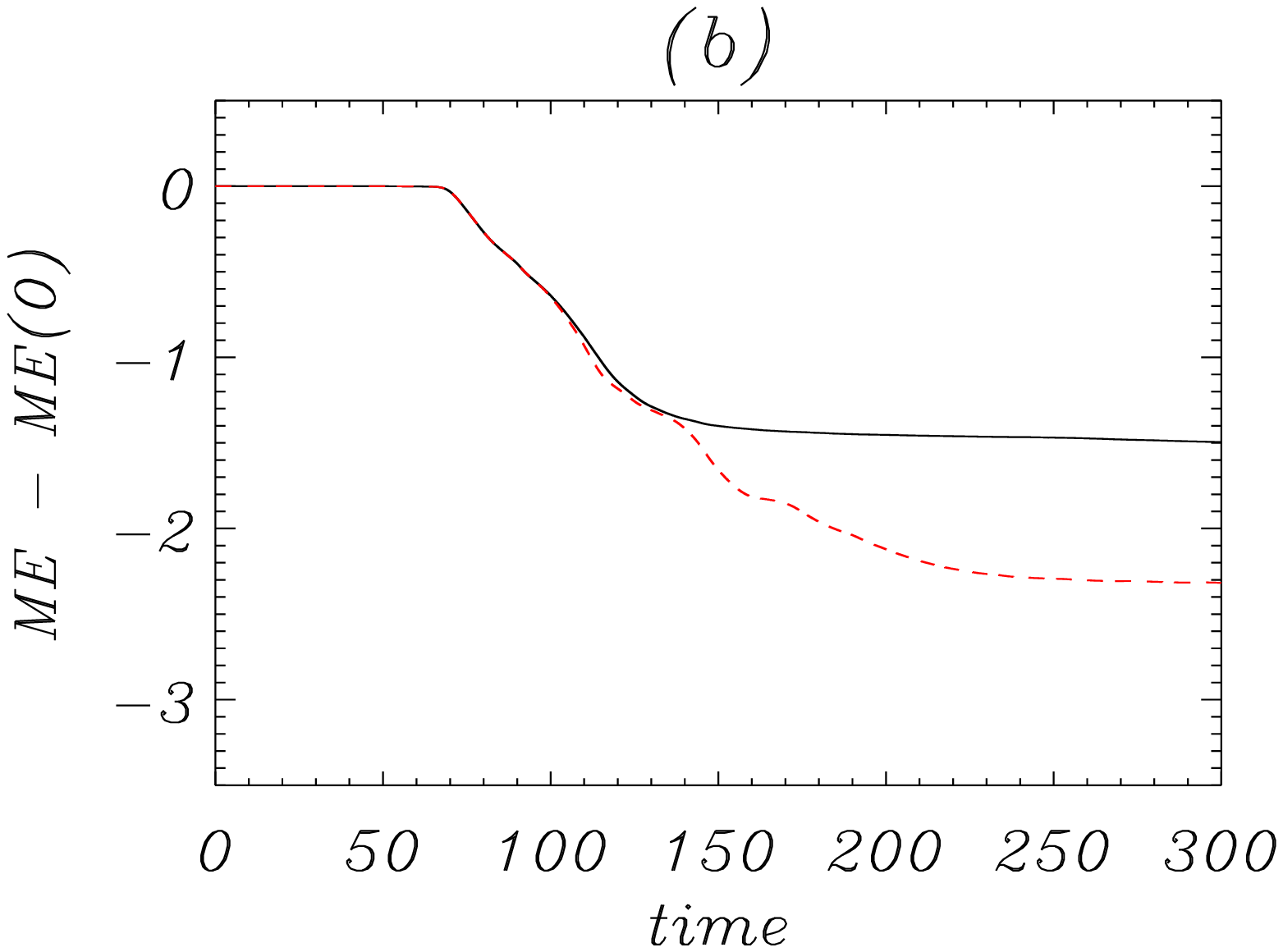}}
{\includegraphics[width=0.49\textwidth]{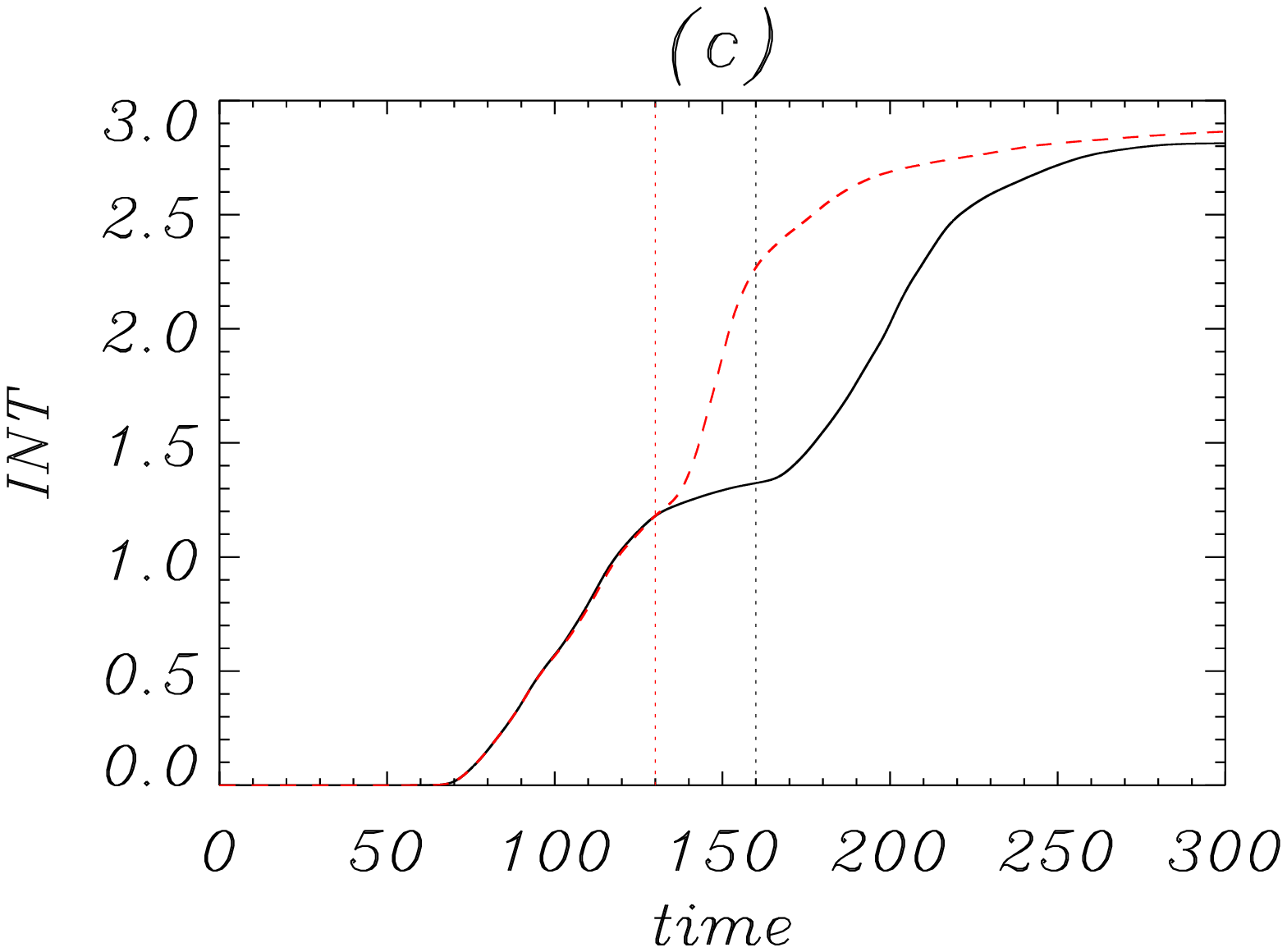}}
{\includegraphics[width=0.49\textwidth]{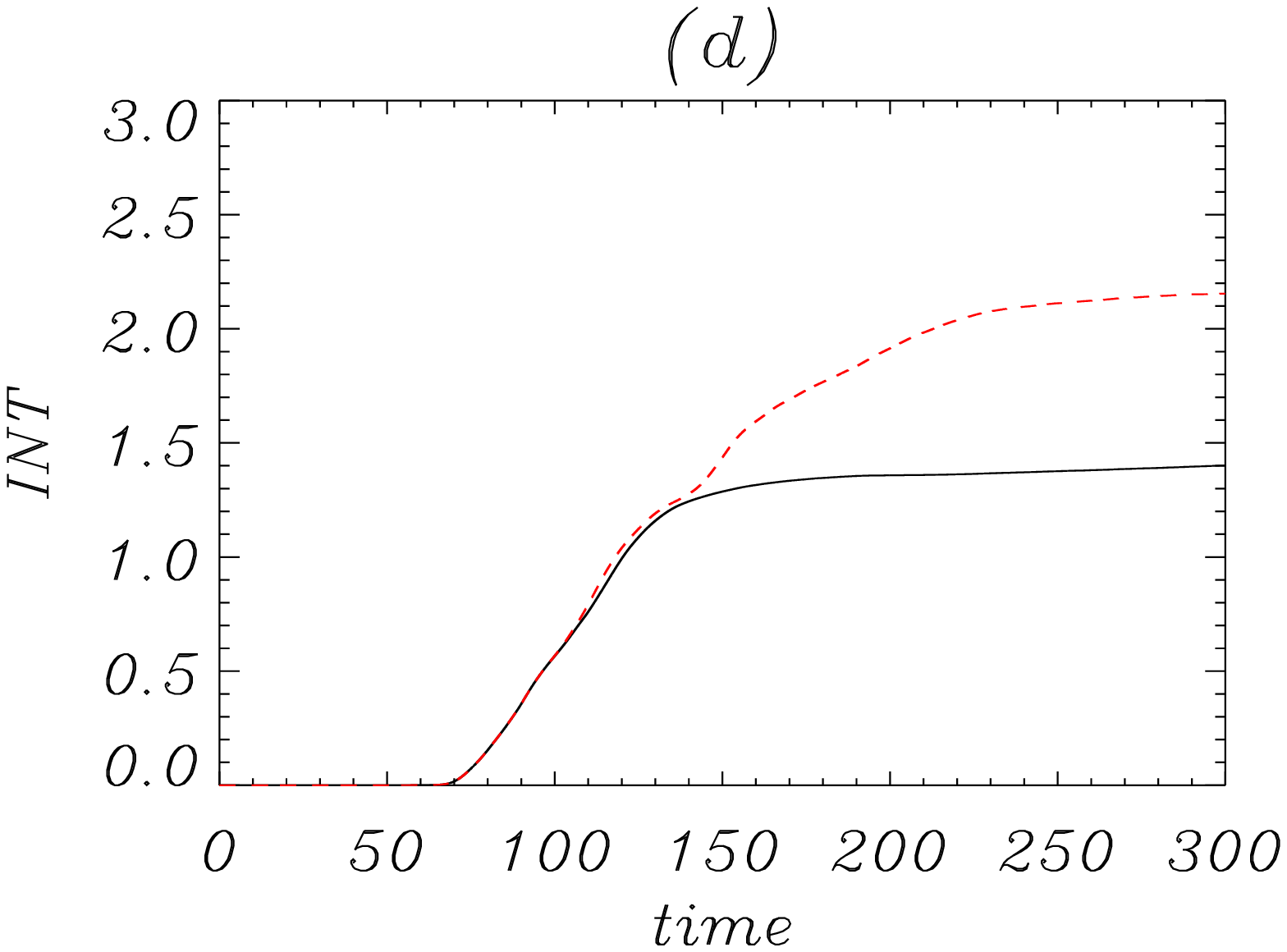}}
\caption{Top row shows the total magnetic energy (ME) minus the initial value as a function of time for \textbf{(a)} Case 1 (\textit{black solid curve}) and Case 2 (\textit{red dashed curve}), 
\textbf{(b)} Case 3 (\textit{black solid curve}) and Case 4 (\textit{red dashed curve}). The total magnetic energy in the volume at time $t=0$ is 175.52
for Case 1 and Case 2 and 174.52 for Case 3 and Case 4. The bottom row shows the temporal evolution of the total internal energy (INT) for \textbf{(c)} Case 1 (\textit{black solid curve}) and Case 2 (\textit{red dashed curve}), 
\textbf{(d)} Case 3 (\textit{black solid curve}) and Case 4 (\textit{red dashed curve}).}
\label{case1_engfig}
\end{figure*}

\begin{figure*}
\centering
{\includegraphics[width=0.45\textwidth]{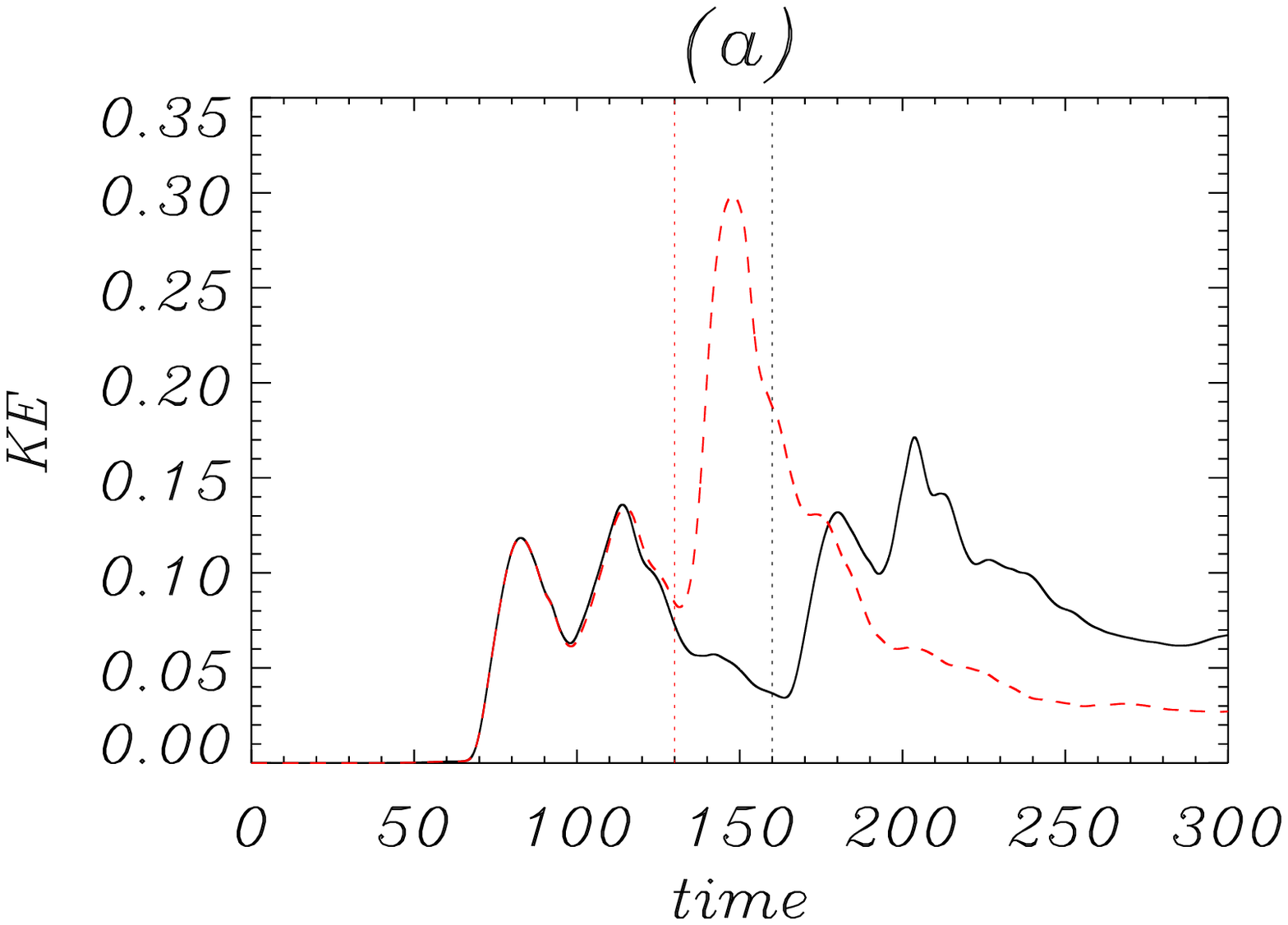}}
{\includegraphics[width=0.45\textwidth]{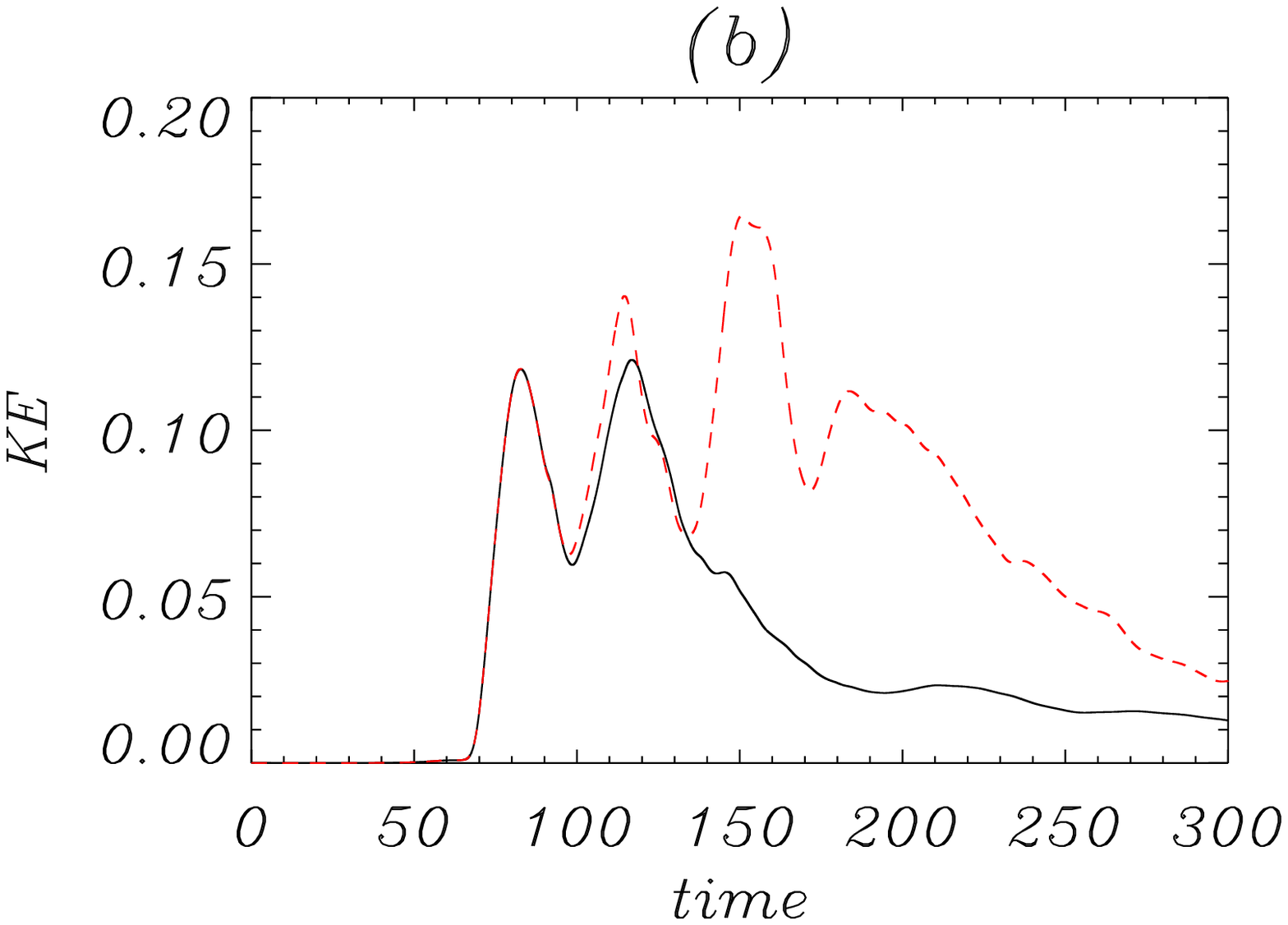}}
\caption{Temporal evolution of the total kinetic energy for \textbf{(a)} Case 1 (\textit{black solid curve}) and Case 2 (\textit{red dashed curve}) and \textbf{(b)} Case 3 (\textit{black solid curve}) and Case 4 (\textit{red dashed curve}).}
\label{case3_engfig}
\end{figure*}

Figures~\ref{case1_engfig} and \ref{case3_engfig} show the volume integrated magnetic, internal and kinetic energies as functions of time for Cases 1 to 4. 
In each case, the rapid
reduction in magnetic energy starts during the development of the kink instability and the onset of magnetic reconnection. 
We discuss the evolution of each form of energy in turn.

\begin{figure*}
\centering
{\includegraphics[width=0.49\textwidth]{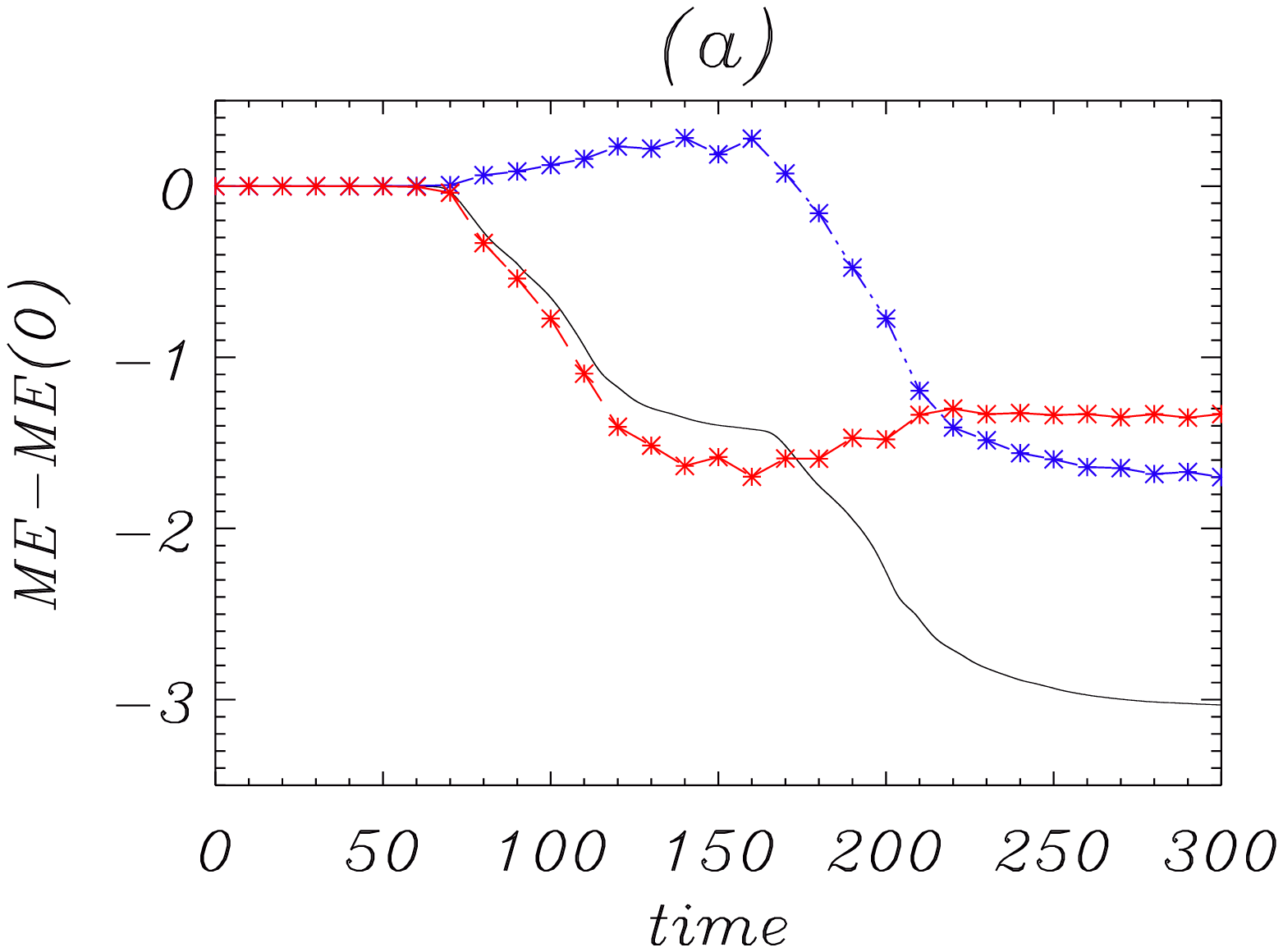}}
{\includegraphics[width=0.49\textwidth]{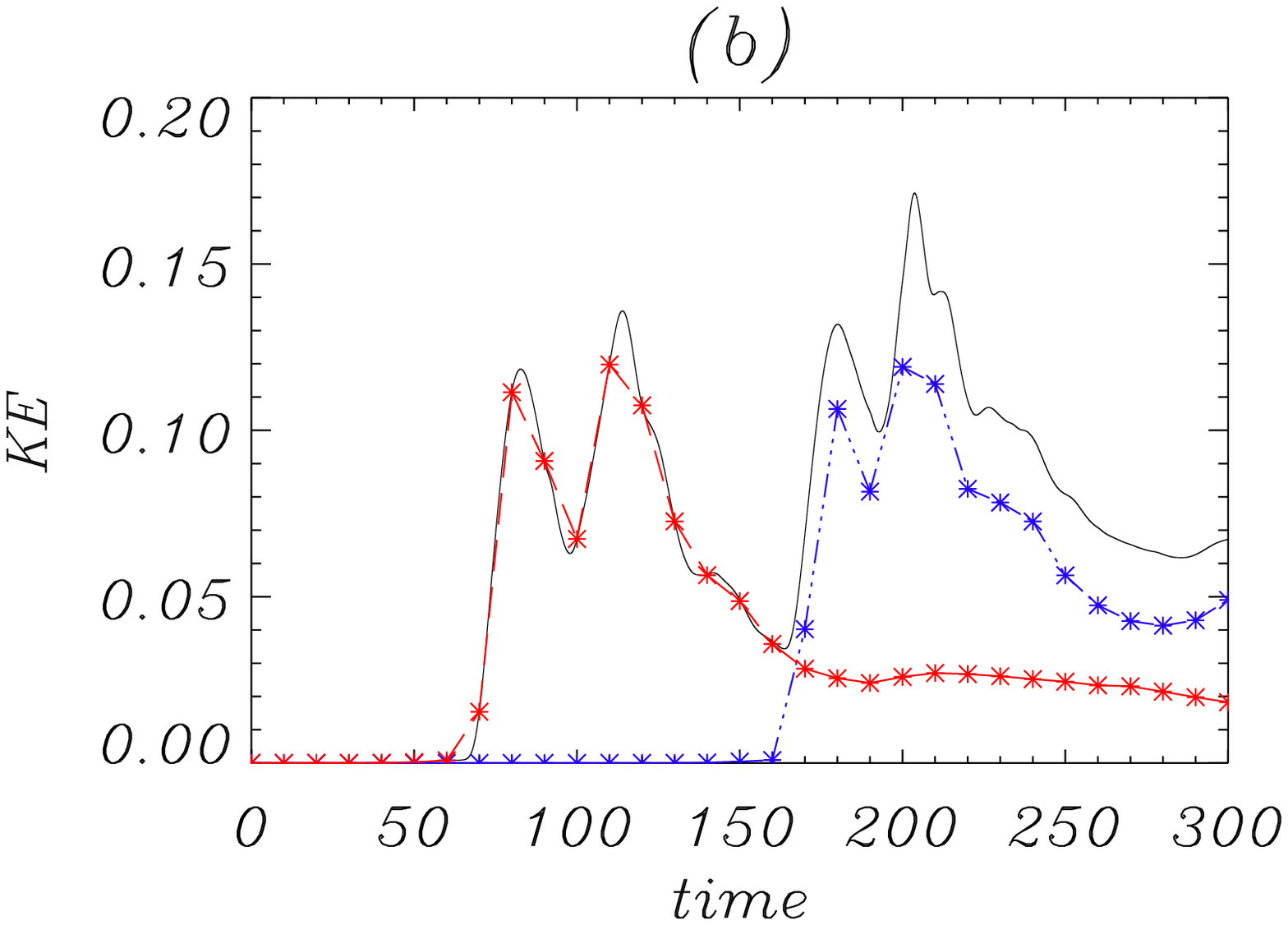}}
{\includegraphics[width=0.49\textwidth]{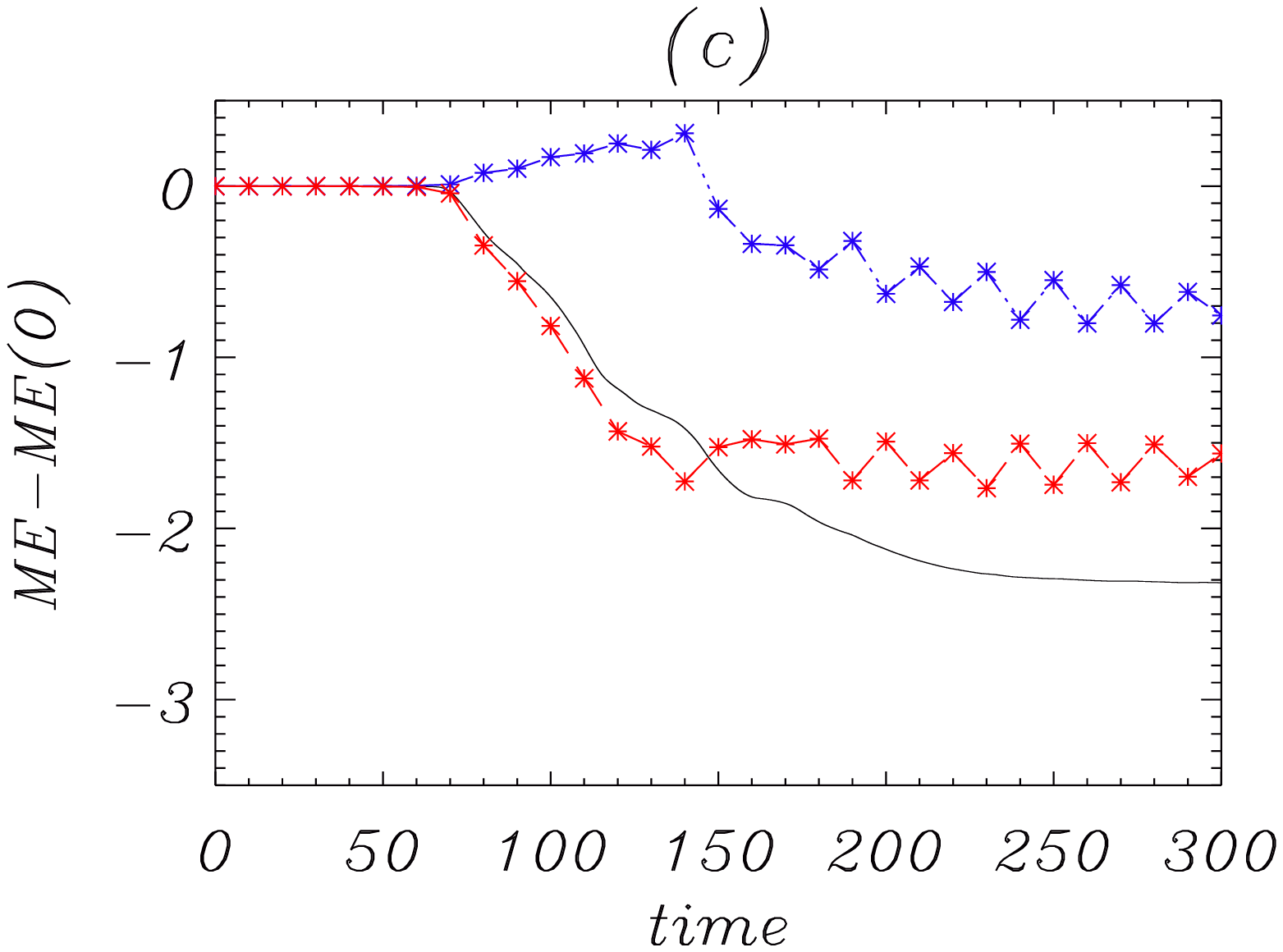}}
{\includegraphics[width=0.49\textwidth]{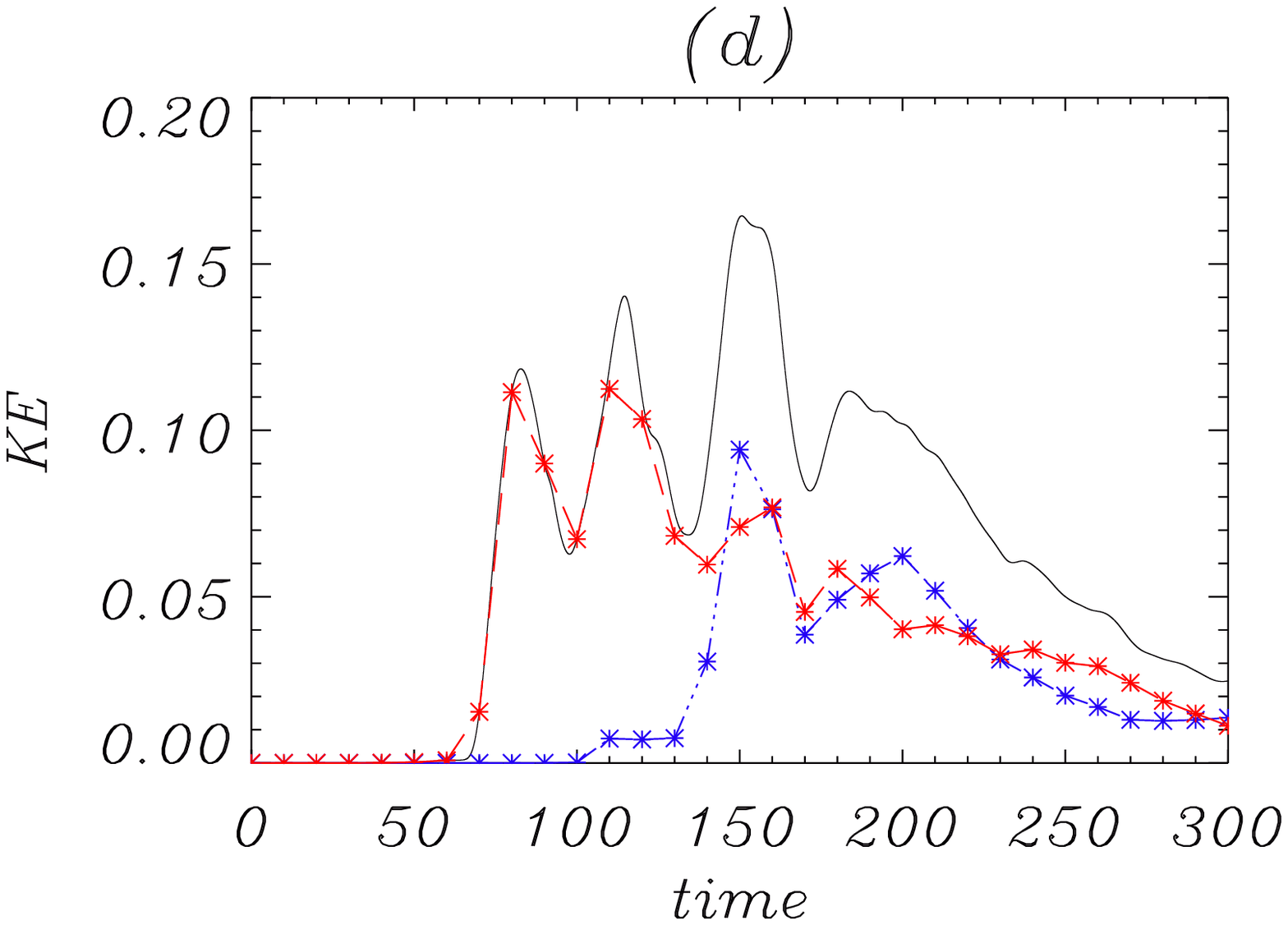}}
\caption{Temporal evolution of the volume integrated energy plots of \textbf{(a)} the change in magnetic energy and \textbf{(b)} the kinetic energy for Case 1 and  
\textbf{(c)} the change in magnetic energy and \textbf{(d)} the kinetic energy for Case 4. 
The magnetic energy at time $t=0$ is 175.52 for Case 1 and 174.52 for Case 4. The black curves show the energy profiles as in Figure~\ref{case1_engfig}, while the red curves show the 
energy profiles on the left-hand side of the domain and the blue curves for the left-hand side.}
\label{case1_en_count}
\end{figure*}

\subsubsection{The time evolution of magnetic, internal and kinetic energies}
Figure~\ref{case1_engfig}a (\textit{black solid curve}) shows the change of total magnetic energy over the volume in Case 1. It stays at its initial value until $t=65$, after which there is a release of magnetic energy. The loss
of magnetic energy becomes significantly slower between $t=125$ and $t=165$. 
This continued, but now slower, decrease in magnetic energy, is an indication that the magnetic field is relaxing towards a minimum state, as discussed by \cite{taylor1974,taylor1986}. 
This behaviour repeats once again from $t=165$ onwards, as the instability of the second thread is triggered. 
The total magnetic energy has dropped by about 3.03 by the end of simulation. 

The \textit{red dashed curve} in the same plot shows the magnetic energy evolution of Case 2. It follows the same initial evolution as Case 1. 
However, the second phase of magnetic energy release occurs slightly earlier, at time $t=135$, as the threads are now placed right next to each other. 
The closer the threads are together the sooner the second instability is triggered. The time difference can be interpreted as a combination of the time for a fast wave to travel and that the amplitude of the wave will be smaller (by about a factor of
2). Hence, the instability needs longer to reach the same amplitude.

By the end of the simulation in Case 2, the magnetic energy has decreased by 3.07. When magnetic energy is released, the internal energy shows a similar size of increase at
the same times. Hence, the majority of the released magnetic energy goes into a rise in the internal energy of the plasma. 
Note that the overall energy released in Cases 1 and 2 is almost the same. If the threads are unstable, then they will relax to their lowest energy state, regardless of the specific dynamical evolution. It is only a question
of when the second thread is destabilised. However, Case 1 releases slightly less energy, indicative of the incomplete relaxation (the energy of two weakly twisted threads being slightly higher than a single tube).

For Case 3, we now assume that the left-hand thread is stable with $\lambda = 1.4$. The two threads are placed the same distance apart as in Case 1. As shown by the \textit{black solid curve} in Figure~\ref{case1_engfig}b, 
there is no loss of magnetic energy
until the kink instability begins to dominate (at $t=65$). It then begins to decrease, as in Case 1, but this time there is no second energy release. The magnetic energy release becomes 
much gentler from $t=130$ onwards
as the thread continues to relax towards its minimum energy state. The overall magnetic energy has decreased by just over 1.5 by the end of simulation, which is, about 49\% of the result in Case 1. As one might expect, only the 
unstable magnetic thread has released its stored energy. 

The interesting result is that more magnetic energy is released if these threads are moved closer together, as in Case 4. The \textit{red dashed curve}s in Figures~\ref{case1_engfig}b and \ref{case1_engfig}d show the evolution of the energies. 
At time around $t=138$, the magnetic energy begins to decrease again, similar to Case 2. Hence, the disturbance of the first thread, due to the kink instability, has resulted in the release of magnetic energy in the
nearby stable thread. The total energy released is 2.3, approximately 0.7 less than in Cases 1 and 2. However, the initial magnetic energy for Cases 3 and 4 is approximately 1 unit smaller than Cases 1 and 2, due to the weaker twist and
smaller axial field strength in the 
left-hand thread. 

The magnitude of the total kinetic energy is significantly smaller than the magnitude of the total internal energy. Hence, most of the magnetic energy released is appearing as heat and not motion.
However, the small changes in the kinetic energy are easier to detect and rapid increases in
kinetic energy are clear indicators of dynamical events. In all four cases, the kinetic energy only begins to rise rapidly at about $t=65$ (see Figure~\ref{case3_engfig}) as the unstable 
right-hand thread is excited. It then peaks twice, reaching a maximum of 0.136 around $t=115$, followed by a slow decay. Without the second thread, this slow decay
would continue as the magnetic field relaxes. For Case 1, the kinetic energy begin to rise again around $t=165$, as the second thread is driven unstable. Case 2, however, shows a much earlier rapid rise at time 
$t=132$, followed by a fast decay around $t=150$. By the end of the simulations, there is more kinetic energy left in Case 1, suggesting that this configuration still has to reach its final Taylor relaxed state.

When the left-hand thread has a stable twist profile, the kinetic energy behaves initially as expected, in response to the unstable right-hand thread. The black curve in Figure~\ref{case3_engfig}b peaks only twice, followed by a 
gentle decay without any further increase. Clearly only the energy in the unstable thread has been released. However, in Case 4, the threads are placed next to each other and the kinetic energy shows a dramatic increase 
as the second stable thread is destabilised by the disturbances of the unstable thread.

\subsubsection{The time evolution of the energy in each thread}
We calculate the temporal evolution of the volume integrated energies in the left-hand ($x < -1$) and right-hand ($x > -1$) sections of the plasma volume separately. 
In Case 1, the right-hand side is excited by an initial perturbation, while the left-hand side is only destabilised by the instability in the right-hand side. 
In Figure~\ref{case1_en_count}, the \textit{red curves} show the energy profiles of the right-hand side volume, the \textit{blue curves} the left-hand side volume and
the \textit{black curves} the sum of the two curves. In Figure~\ref{case1_en_count}a the \textit{black curve} shows the two stage 
release of magnetic energy discussed above. The \textit{red dashed curve} shows that the right-hand side actually loses more magnetic energy than the total value, between $t=65$ and $t=180$. Since the majority of this
released magnetic energy goes into internal energy, there is an increase in the pressure in the right-hand volume. This creates an expansion that pushes the plasma into the left-hand volume, compressing 
the magnetic field and increasing the magnetic energy there, as shown in the \textit{blue triple dotted-dashed curve}. The increase in magnetic energy in the left-hand volume reaches a maximum value at the time the energy released 
in the right-hand volume is at its smallest value. However, once the kink instability is triggered in the left-hand volume the magnetic energy is again reduced. The comparable behaviour for the kinetic energy for Case 1 is shown in 
Figure~\ref{case1_en_count}b. 

Figure~\ref{case1_en_count}c shows the time evolution of the magnetic energy for Case 4. Again we see the increase in magnetic energy in the left-hand volume as the energy in the right-hand volume decreases. From $t=140$ the magnetic energy
in the left-hand side starts to decrease. There remains some oscillations, that are not resolved in the time storage of our data. However, the sum of the red and blue curves is smooth, suggesting that this is simply an oscillatory transfer of magnetic energy between 
the two regions. The kinetic energy curves in Figure~\ref{case1_en_count}d follow a similar behaviour to Case 1.

\subsection{Current and  magnetic field line time evolution}
We begin this section by considering the temporal evolution of the magnitude of the current density in Cases 3 and 4 before discussing the structure of the magnetic fields. 

\begin{figure*}[ht]
 \centering
  \includegraphics[width=1.0\textwidth]{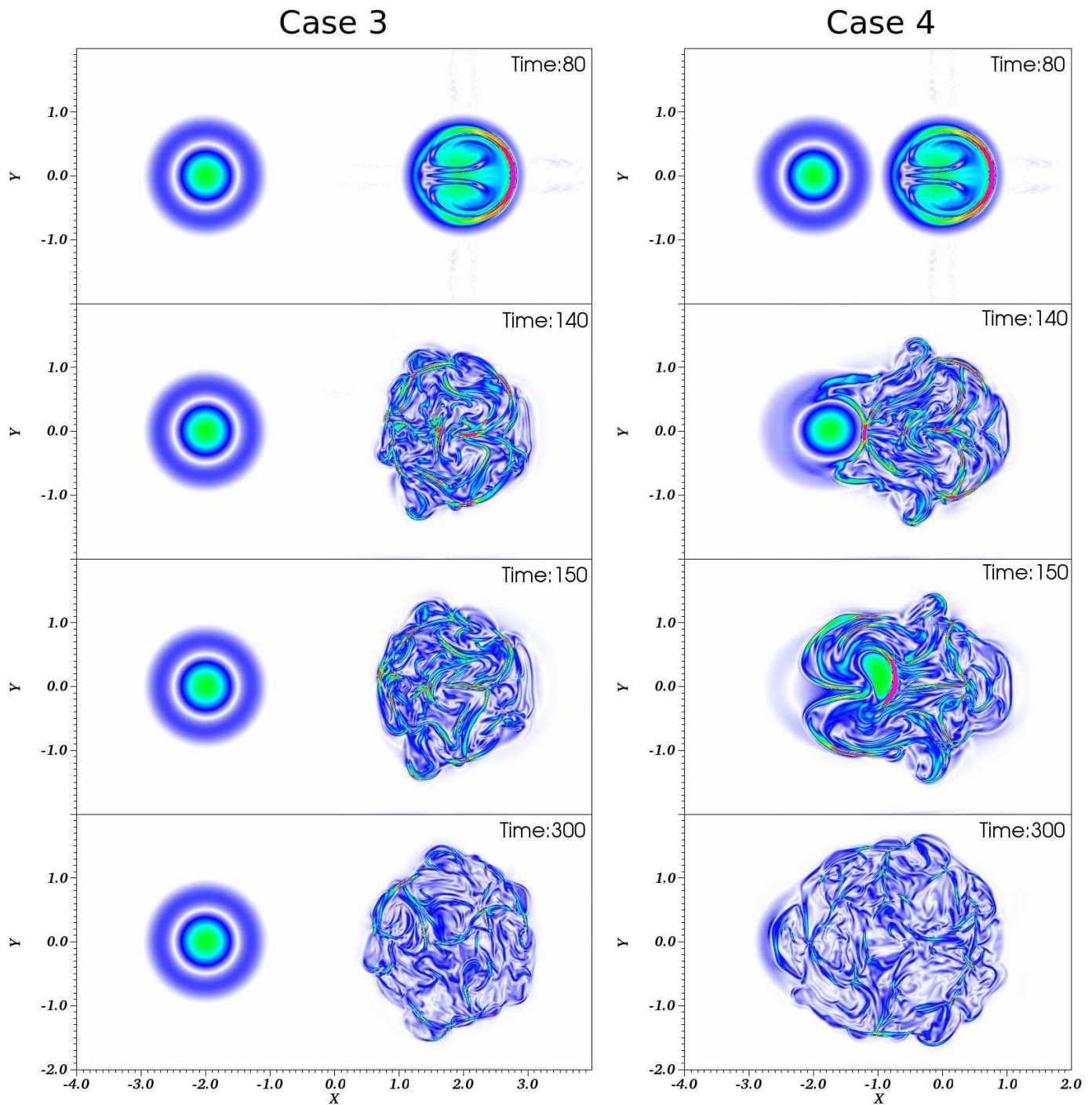}
  \caption{Contour plots of current density magnitude, $j(x,y,0)$, as functions of $x$ and $y$ at $z=0$ at the times indicated. The figures on the left and right are for Case 3 and Case  4
  respectively. The colour scale for the current density goes from 0 (white) to 5 (purple).}
  \label{case1_cd}
\end{figure*}

Figure~\ref{case1_cd} shows the current density magnitude. Due to the initial perturbation, 
a kink instability creates an initial current sheet (red colour in contour plot) in the right-hand magnetic thread at time $t=80$, suggesting that magnetic reconnection is occurring, and this is discussed below. The current within the right-hand thread begins to
fragment and many small scale current structures are seen in the right-hand thread at $t=140$. This has been mentioned by various authors (e.g. Hood et al, 2008) and is a key part of the Taylor relaxation process. 
For Case 3, the second thread is never destabilised, despite the strong disturbances of the right-hand thread as it relaxes.

For Case 4, the left-hand stable thread is actually destabilised around $t=130$. The expansion of the right-hand thread has caused it to strongly interact with the left-hand thread, as shown at $t=140$. The dominant current component is $j_{z}$ and,
in the outer portion of each thread this is negative. Hence, the expansion of the thread means that the touching parts of the two threads mixes the same sign of current. 
If the left-hand thread had the opposite twist, so that the outer part of $j_{z}$ was positive, then the interaction would be significantly reduced and the destabilisation would not take place.
The left-hand thread is strongly disrupted by $t=150$ and now relaxes towards its lowest energy state and, by the final time at $t=300$, there is a clear indication that the two threads
have combined to form a single larger thread, as also shown in Figure~\ref{case34barrows}.

\begin{figure*}[ht]
 \centering
  \includegraphics[width=0.9\textwidth]{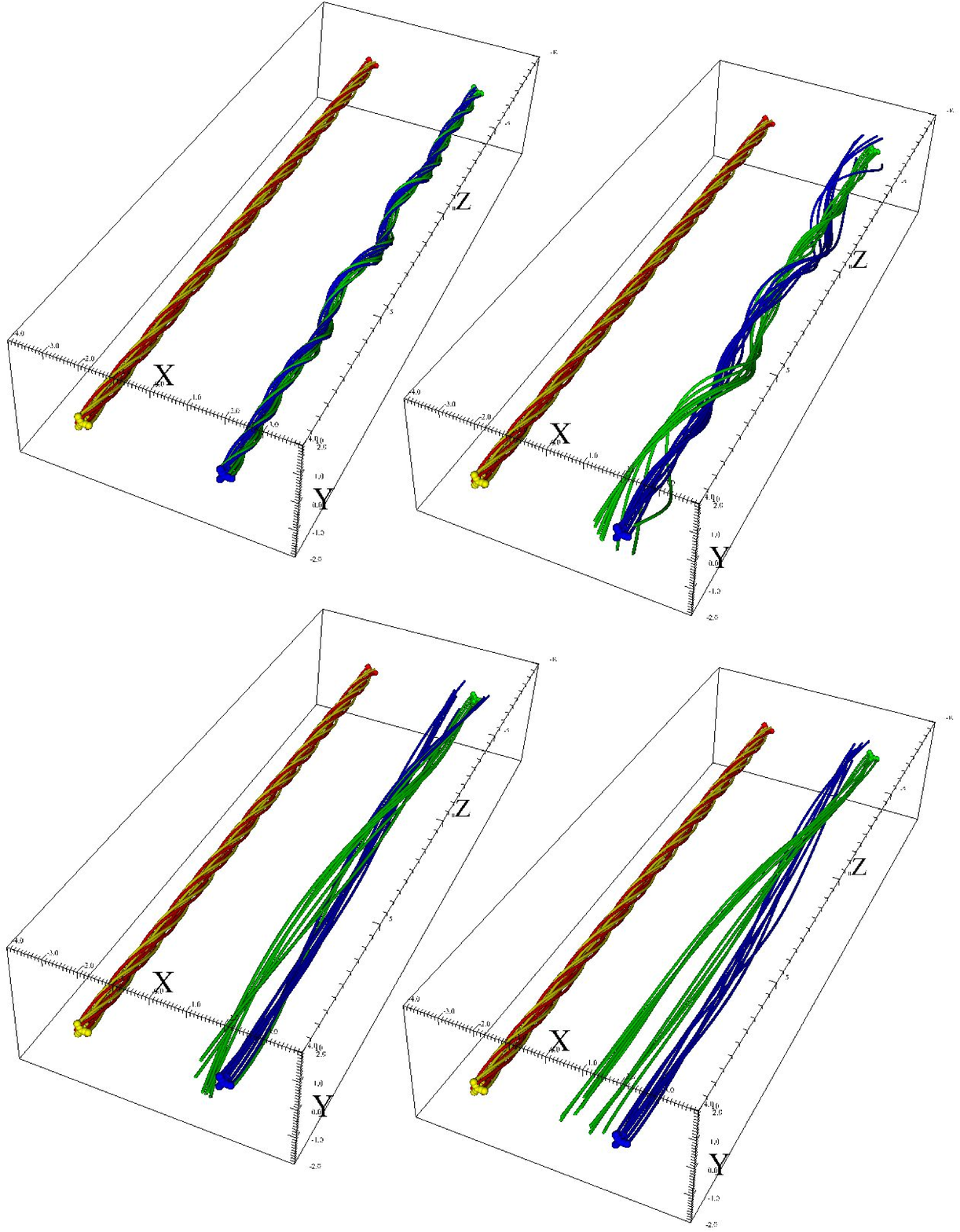}
\caption{Case 3: the field line plots at time: \textbf{(a)} $t=60$ (\textit{top left}), \textbf{(b)} $t=80$ (\textit{top right}), \textbf{(c)} $t=160$ (\textit{bottom left}) and \textbf{(d)} $t=300$ (\textit{bottom right}). The yellow and red field lines are drawn from $(x,y,z)=(-2,0,10)$ and $(x,y,z)=(-2,0,-10)$ while the blue and green field lines are drawn form $(x,y,z)=(2,0,10)$ and $(x,y,z)=(2,0,-10)$.}
  \label{case1_fl}
\end{figure*}

To investigate evidence for reconnection, we track the evolution of the magnetic field lines for Cases 3 and 4. 
The field lines around the centre of each thread are traced from one photospheric end, where the velocity is zero, 
to the other end as shown in Figure~\ref{case1_fl} for Case 3. These are coloured red and yellow for the left-hand thread and blue and green
for the right-hand thread. If there is no reconnection, then the red/yellow and blue/green field lines will lie on top of each other. If there is reconnection, then the ends of the field lines will not pass through
the appropriate footpoint.

At time $t=60$, a helical structure can be seen on the right-hand thread, as the kink instability develops. However, the field lines in the right-hand thread are still lying on top of each other and so
there is no evidence of reconnection at this time. 
At time $t=80$, the field lines are seen to unwind or straighten, giving evidence of reconnection in Figure~\ref{case1_fl}b (top right). In particular, the green field lines start from the thread axis 
at the far end of the right-hand thread. However, these
field lines completely encircle the thread axis at the near end. The green and blue field lines do not follow the same paths. For case 3, there is no reconnection in the left-hand thread.

\begin{figure*}[ht]
 \centering
  \includegraphics[width=0.9\textwidth]{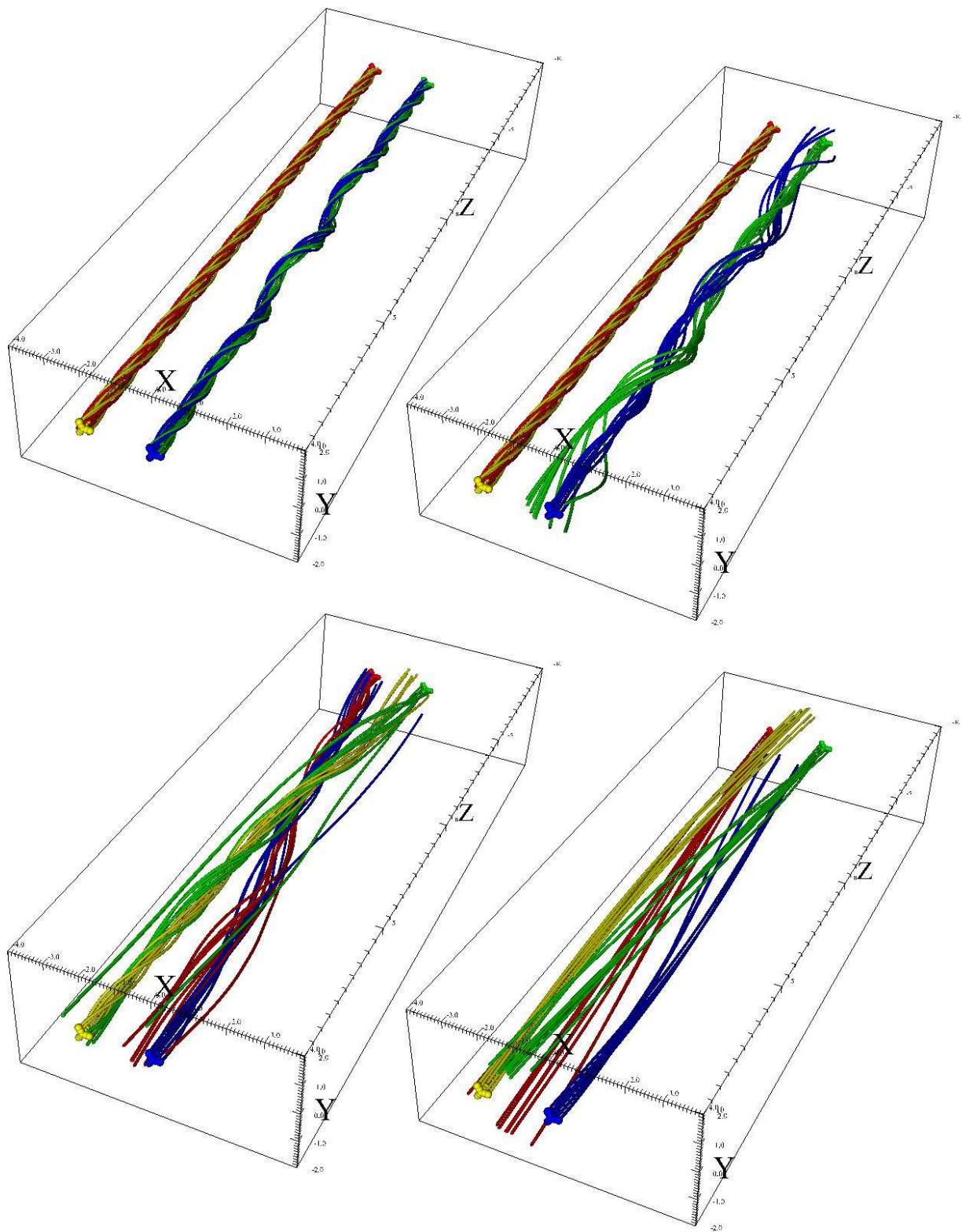}
  \caption{Case 4: the field line plots at time: \textbf{(a)} $t=60$ (\textit{top left}), \textbf{(b)} $t=80$ (\textit{top right}), \textbf{(c)} $t=160$ (\textit{bottom left}) and \textbf{(d)} $t=300$ (\textit{bottom right}). The yellow and red field lines are drawn from $(x,y,z)=(-2,0,10)$ and $(x,y,z)=(-2,0,-10)$ while the blue and green field lines are drawn form $(x,y,z)=(0,0,10)$ and $(x,y,z)=(0,0,-10)$.}
  \label{case2_fl}
\end{figure*}

The evolution of the field lines for Case 4 is shown in Figure~\ref{case2_fl}. The initial behaviour is exactly the same as Case 3 for $t=60$ and $t=80$. However, once the second instability is triggered ($t=160$), the field lines
of the left-hand thread also begin to unwind. By the end of simulation at $t=300$, what is clear is that the various coloured field lines are now wrapping around each other, forming one
weakly twisted magnetic thread that has a rotation of about $90^\circ$ about a common axis. It will be interesting to see if future simulations can determine (i) how far apart the individual threads must be before this destabilising of a stable thread occurs and
(ii) whether the threads merge to form one single twisted structure.

\subsection{Heating and temperature structure}
\begin{figure*}[ht]
 \centering
 {\includegraphics[width=0.49\textwidth]{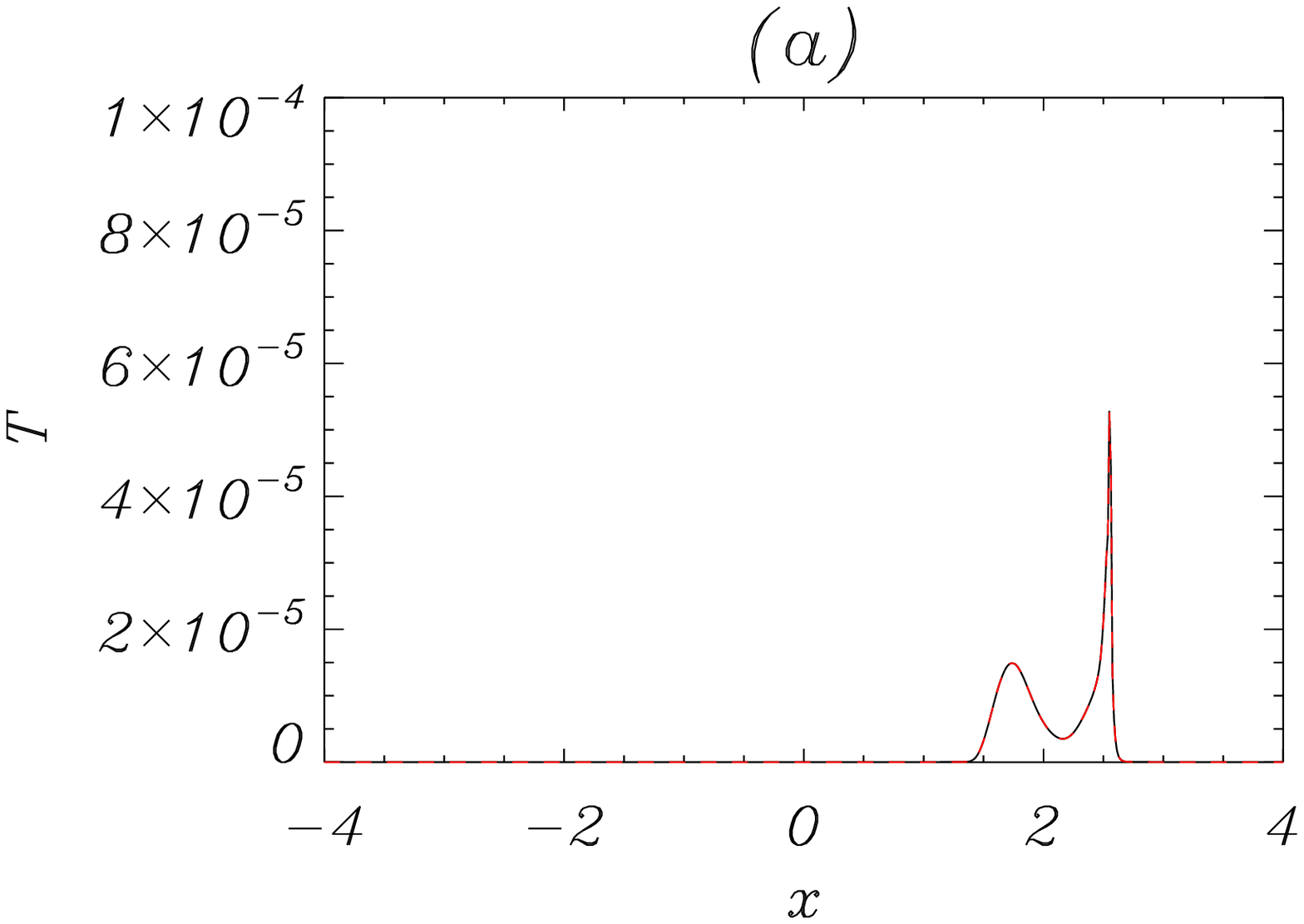}}
 {\includegraphics[width=0.49\textwidth]{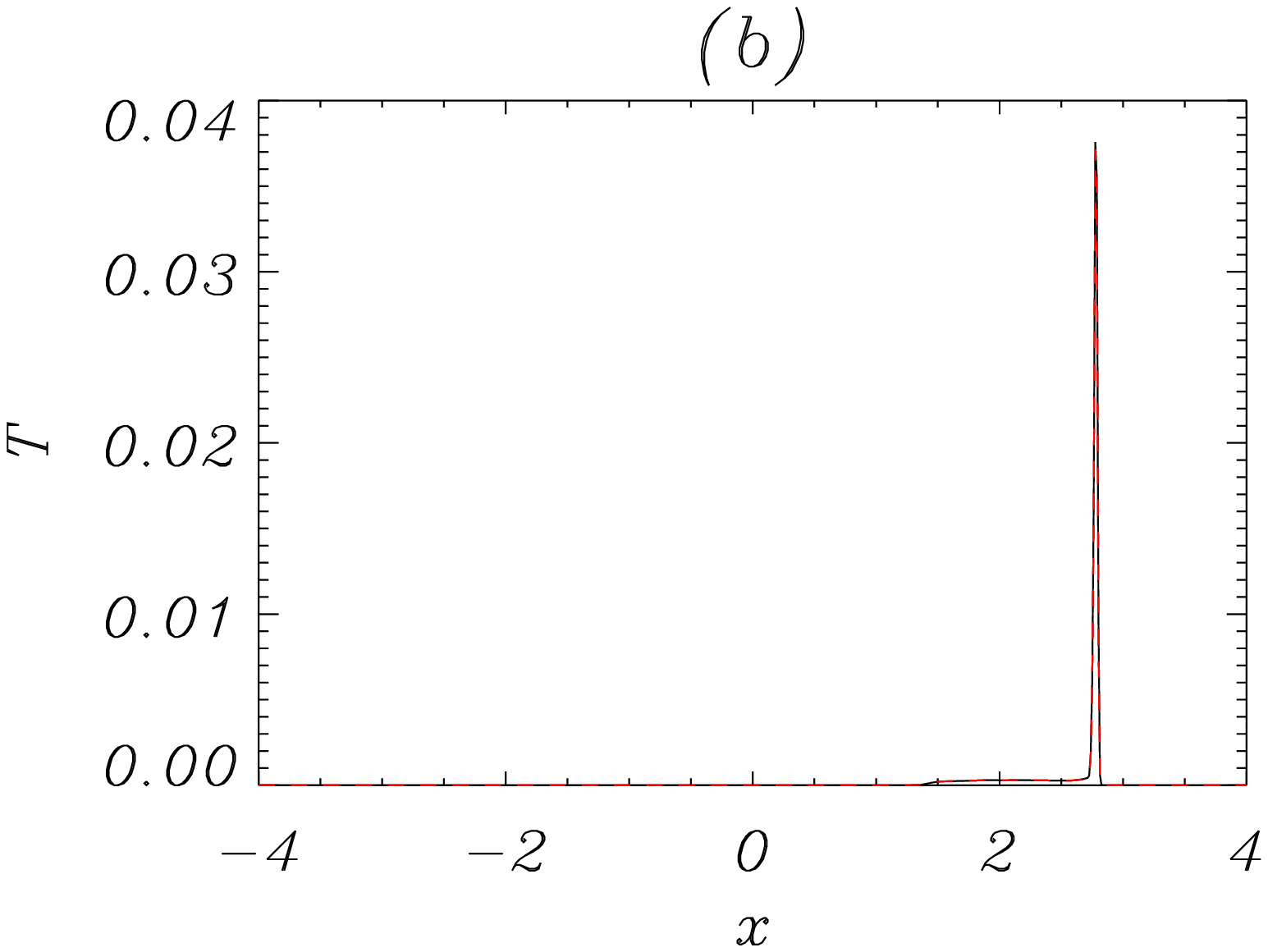}}
 {\includegraphics[width=0.49\textwidth]{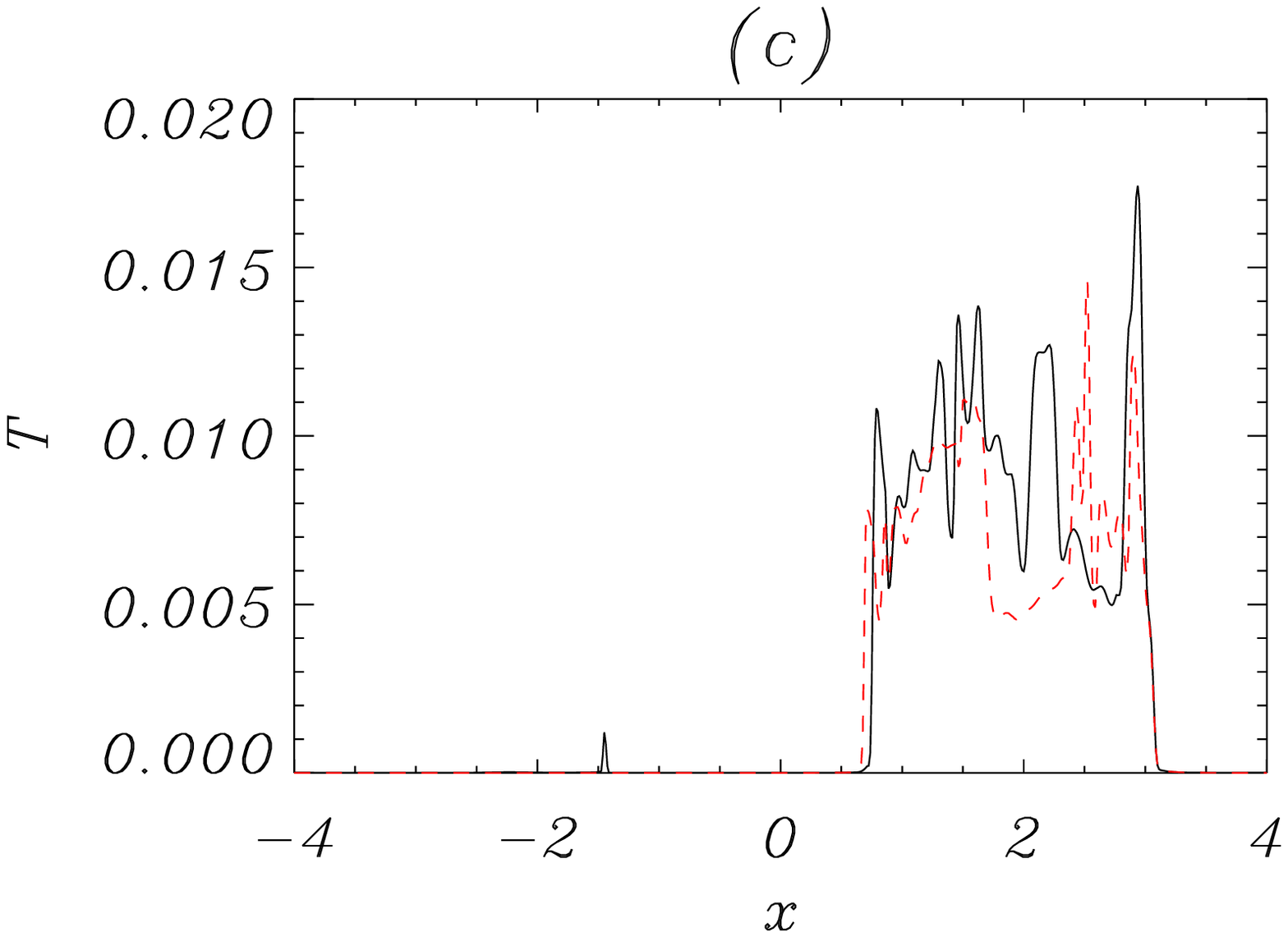}}
 {\includegraphics[width=0.49\textwidth]{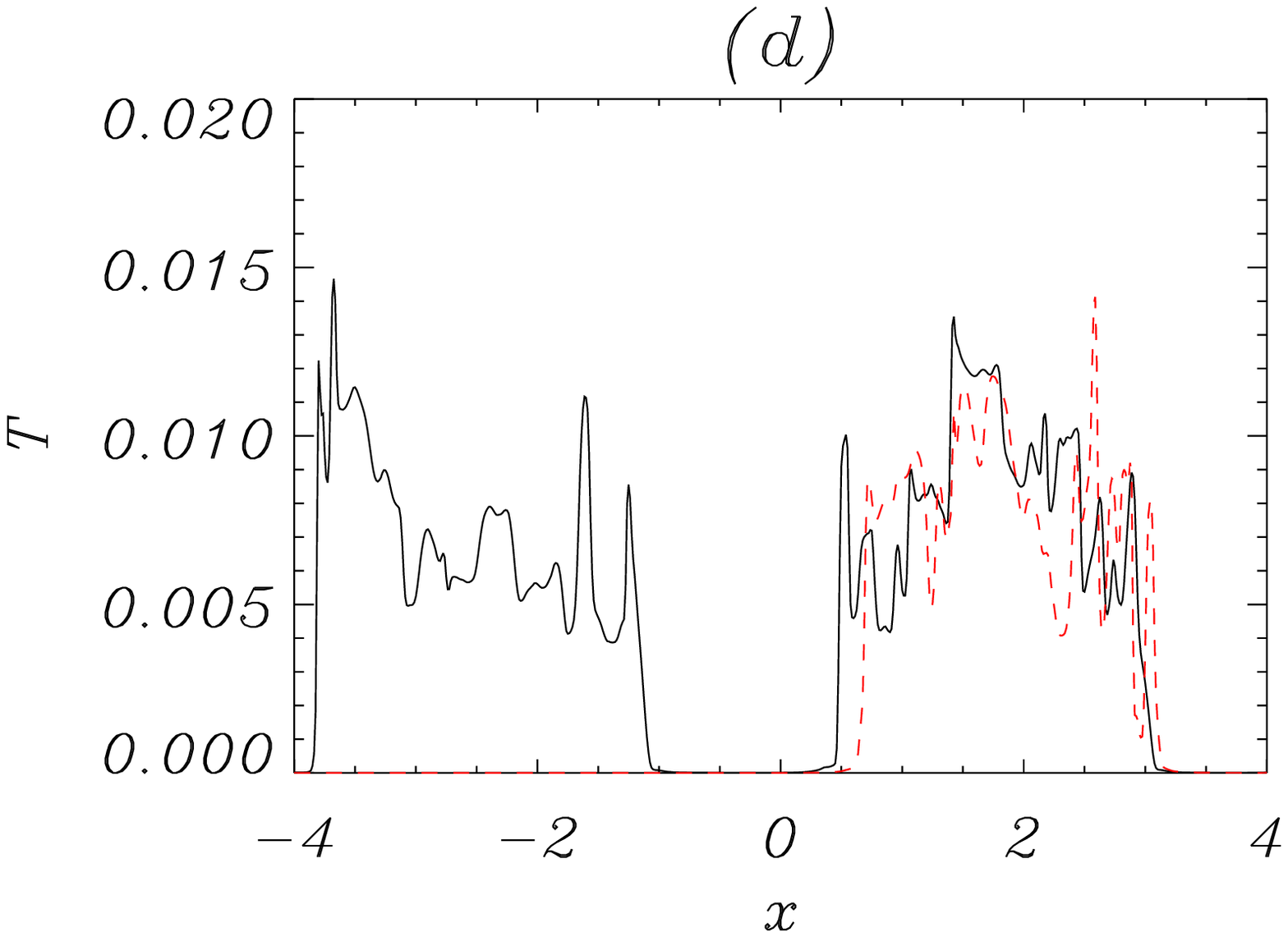}}
  \caption{Temperature plots, $T(x,0,0)$ for Case 1 (\textit{black}) and Case 3 (\textit{red}) at times $t=$ \textbf{(a)} 60, \textbf{(b)} 80, \textbf{(c)} 160 and \textbf{(d)} 300.}
  \label{case1_temp}
\end{figure*}

In this section, we will study how the temperature, in a cut across the mid-plane, behaves for each of the four cases as functions of $x$ on $y=0$ and $z=0$.  
For Case 1 (\textit{black solid curves})
and Case 3 (\textit{red dashed curve}s) and at times $t=60$ and $t=80$, the largest temperatures are where the current sheet is forming (Figure~\ref{case1_temp}). The two curves lie on top of each other.
The peak dimensionless temperature is $0.038$ at $t=80$ and the peak temperature
is clearly confined to the current sheet. For Case 1, a new instability develops within the left-hand thread and the temperature starts to rise in the current sheet at $x=-1.4$. 
As the instability in develops, the temperature within the left-hand side thread becomes comparable to one on the right. Note that the hot temperature in these threads is now spread over a distance of about 3 in each thread, 
namely 1.5 times their original diameters.

\begin{figure*}[ht]
 \centering
 {\includegraphics[width=0.45\textwidth]{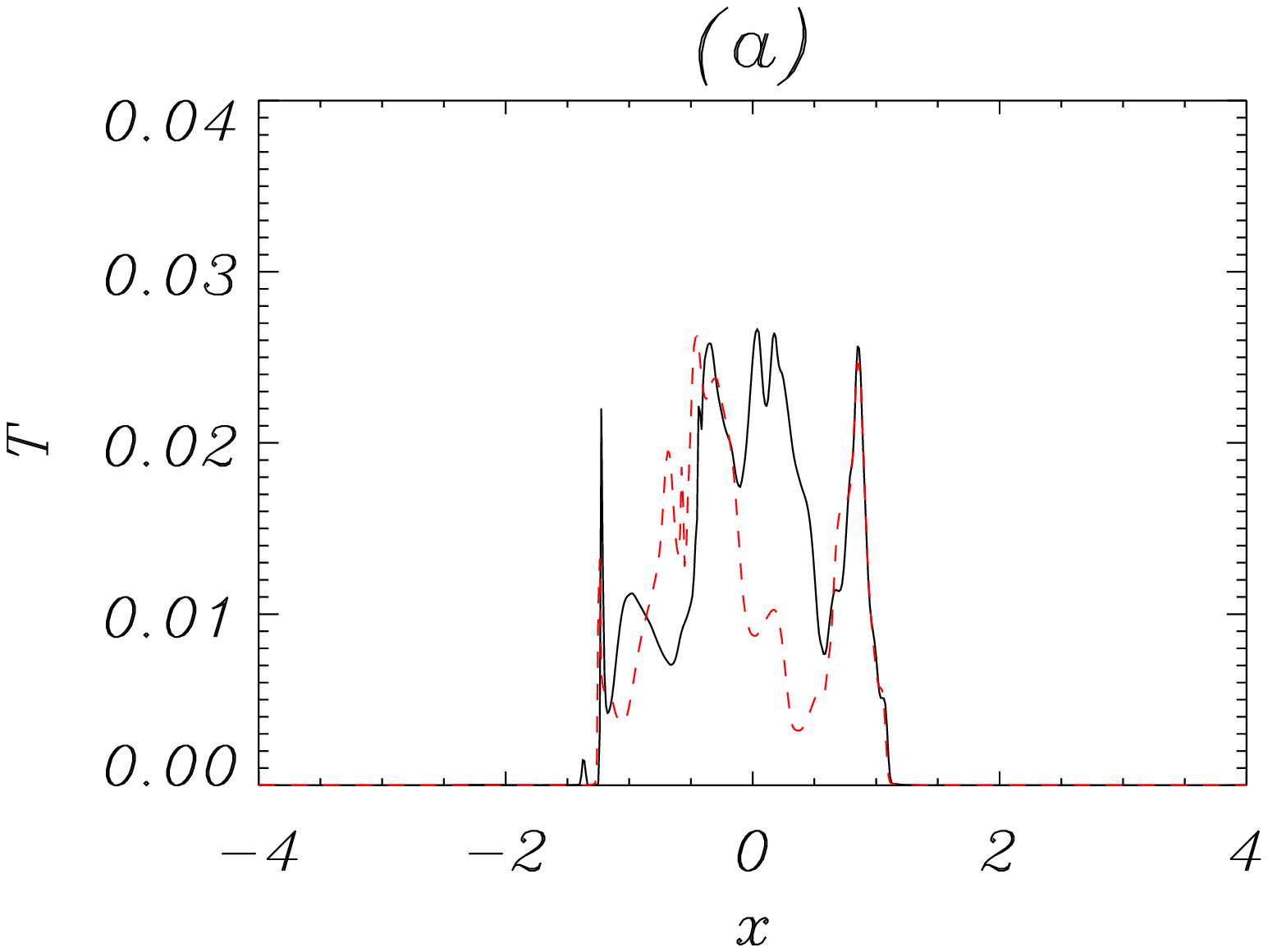}}
 {\includegraphics[width=0.45\textwidth]{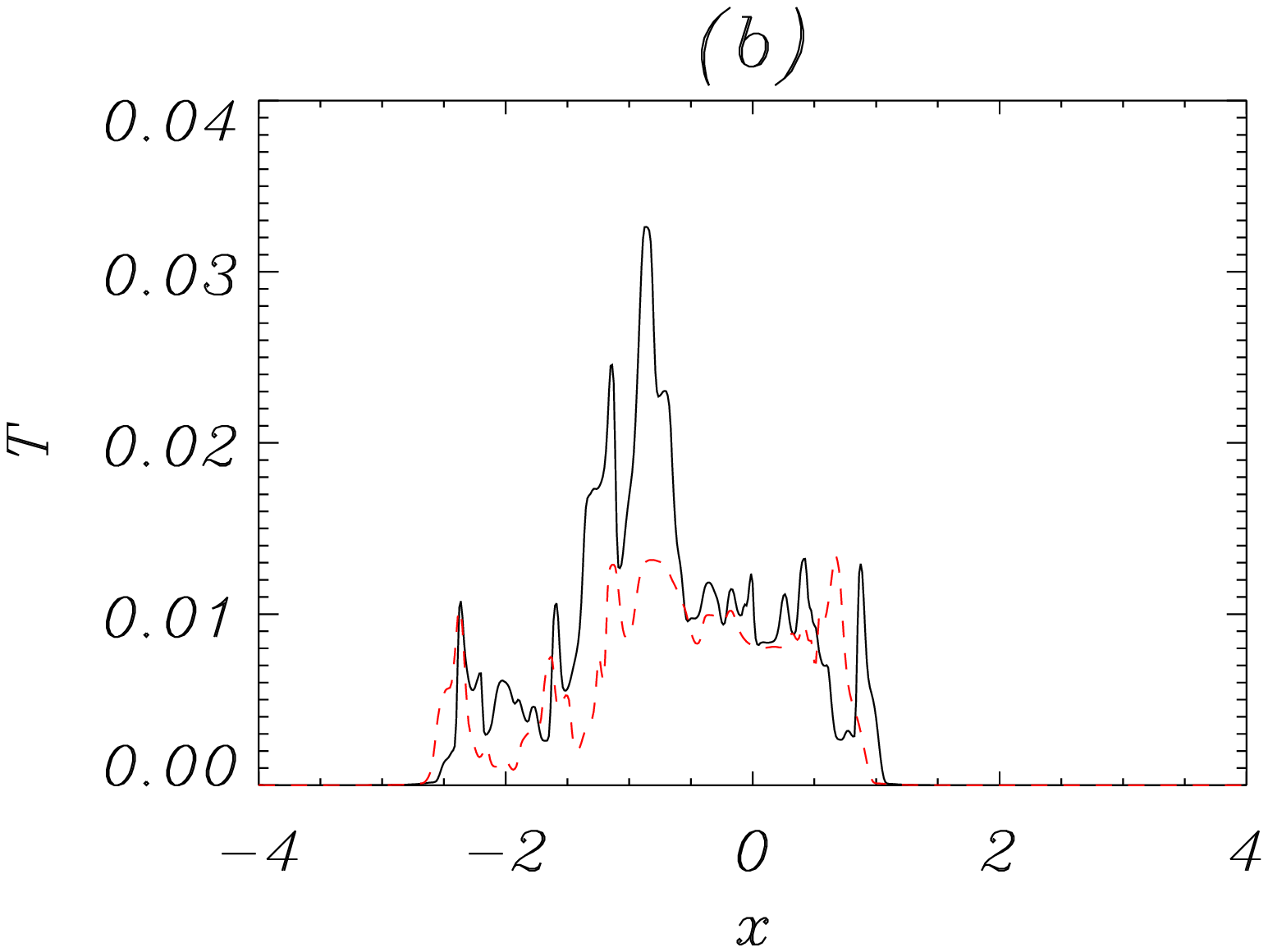}}
  \caption{Temperature plots, $T(x,0,0)$ for Case 2 (\textit{black}) and Case 4 (\textit{red}) at times \textbf{(a)} $t = 130$ and \textbf{(b)} $t = 300$.}
  \label{case2_temp}
\end{figure*}

The temperature profile for Cases 2 (\textit{black solid curves}) and 4 (\textit{red dashed curve}s) are shown in Figure~\ref{case2_temp}. The initial evolution of the temperature curves follows the results of Figure\ref{case1_temp} for
$t=60$ and $t=80$ and are not shown. At $t=130$, a small temperature peak begins to show up near $x=-1.4$ within the left-hand side thread as the second current sheet begins to form. 
The temperature rise is smaller for Case 4 
than for Case 2. In both cases, these threads are merged into one larger thread, from $x=-2.5$ to $x=1.0$ with hot plasma right across this region. The maximum temperature in Figure~\ref{case2_temp}b 
is significantly larger for Case 2 than Case 4, since there is more free magnetic energy available to heat the plasma. What is surprising is that the maximum temperature of Case 2 is actually higher than Case 1. 
So, although both threads are unstable, the temperature is higher when the two unstable threads are touching each other. This is most likely due to the fact that the plasma volume being heated is smaller when the 
two threads combine than when they are separate. A smaller volume for the same energy released will result in a hotter plasma.

It is only Case 3 that has a single magnetic thread heated. In all the other cases, hot plasma is spread across the volume of both threads. 
Hence, since the plasma is optically thin, these magnetic structures will appear brighter than Case 3,
when viewed from the side.

\begin{figure*}[ht]
 \centering
 {\includegraphics[width=0.85\textwidth]{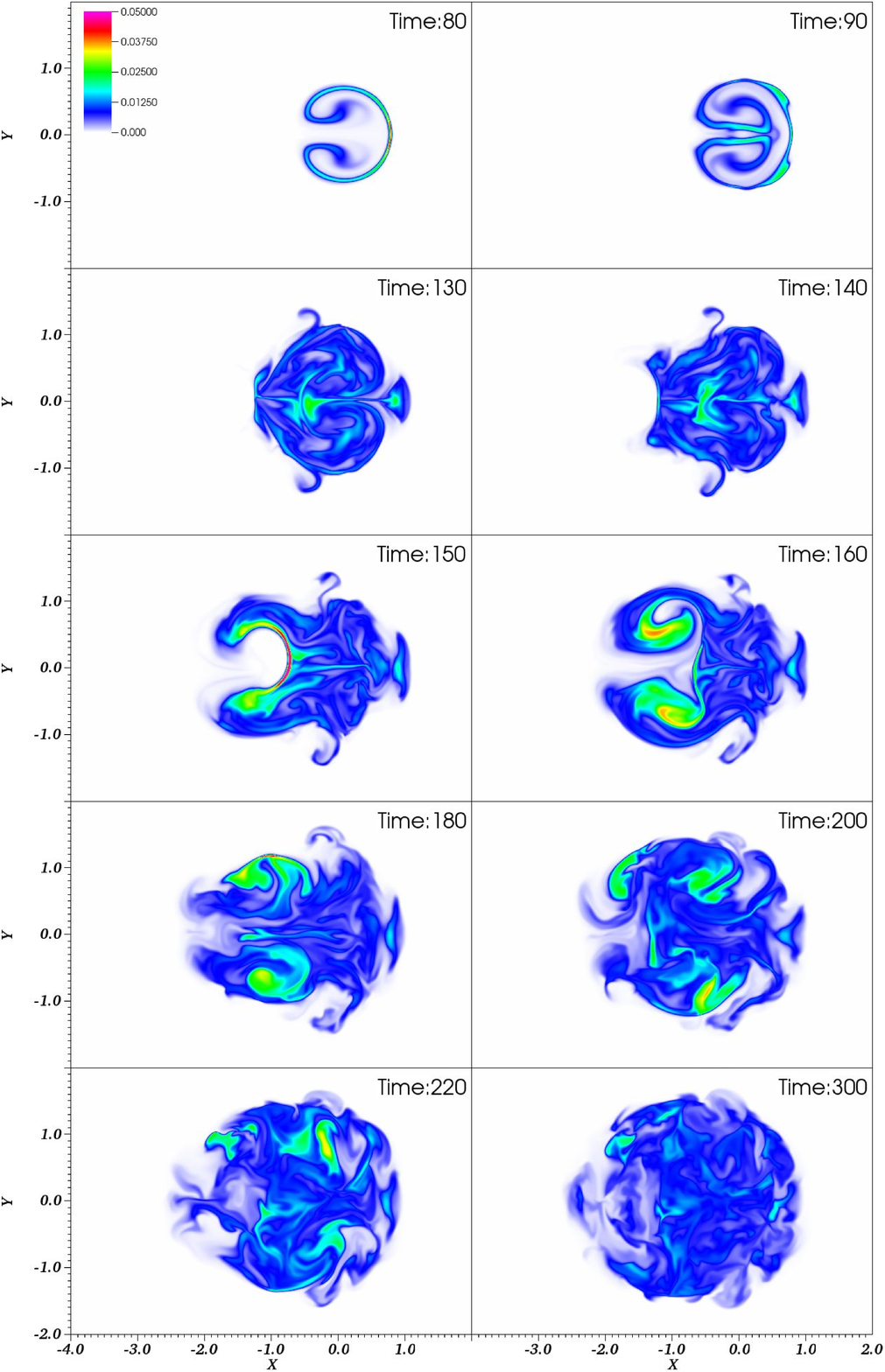}}
  \caption{Temperature at the mid-plane, $T(x,y,0)$, for Case 4 at the times indicated.}
  \label{Case4temp_midplane}
\end{figure*}

The heating in Case 4 is the most interesting, since it is the only case where a single stable thread is destabilised by its unstable neighbour and able to release its free magnetic energy. The the mid-plane temperature, at eight
different times, is shown in Figure~\ref{Case4temp_midplane}. The stable thread is destabilised around $t=140$ and a clear sharp boundary in the temperature is seen on the left-hand side of the heated
plasma. This develops into a hot arc at $t=150$, that is similar is nature to the initial current sheet that forms around $t=80$. By the end of the simulation, a large area has been heated.

The effectiveness of heating of coronal loops through the excitation of reconnection events in multiple magnetic threads can be assessed by the studying the resulting temperatures after the magnetic field
has relaxed. In our simulations, the dimensionless temperature after heating is approximately in the range of $T= 0.005 \sim 0.02$. To convert these to actual coronal values, we consider a typical magnetic field strength 
of $B_0 = 50$ G and a typical mass density of $\rho_0 = 1.67 \times 10^{-13}$ kg m$^{-3}$. The reference temperature is approximately $1.4 \times 10^{10}$K.
Therefore, the dimensional temperature in our simulations is around $T  \sim 7$ to $28 \times 10^7$ K. 
These values are high compared to observed values. Of course, we expect the actual values to be smaller with the inclusion of thermal conduction. This effect could reduce the maximum temperature in the system 
by about a factor of 10 \citep{botha2011}. Optically thin radiation may reduce this even more. However, the aim of this work is not to reproduce exactly the coronal values but to show that an avalanche
can be initiated if the magnetic threads are sufficiently close together.

\section{Discussion and conclusions}\label{sec:discussion}
We have run a series of numerical experiments to investigate the heating of a coronal loop consisting of a number of smaller magnetic threads. In our experiments, we consider two magnetic threads and always have one thread
that is unstable to the kink instability and this thread is excited by an initial velocity perturbation. The second thread is either stable or unstable and it is placed either beside or a distance of 2 units away from the 
edge of the unstable thread.
This gives four different combinations for the arrangement of the magnetic threads. Table~\ref{two_summaries} presents the times at which there is a rapid decrease in magnetic energy in the first thread and the second thread. This
gives a reasonable estimate of when the main heating is beginning in the plasma. For all cases, the first thread begins to release its free energy at the same time. For Case 2, we see that the start of the second energy release occurs
sooner when the threads are closer together than when they are further apart. For Cases 2 and 4, the energy is released at almost the same time, even though the second magnetic thread in Case 4 is stable on its own.

A major result from these simulations is that an individual stable magnetic thread can be destabilised by a neighbouring unstable thread, if they are close enough together and have the same sense of twist. 
This triggering provides the first step in developing MHD avalanche models. {The idea of an avalanche mechanism for energy release in a magnetically complex corona began with the work of \cite{lu1991} and \cite{lu1993}. 
They defined a magnetic field over a three-dimensional grid and imposed \lq rules\rq\  at each point 
that determined whether magnetic energy dissipation occurred as a function of the local field stresses. Dissipation at one point could lead a neighbouring point to become unstable, and so on, hence creating an \lq avalanche\rq . They found that the avalanche sizes 
scaled as $E^{-1.4}$. Subsequent work \citep{vlahos1995,vassiliadis1998} ensured that Maxwell's equations were satisfied within this generic approach, and increasingly sophisticated models have been developed
\citep{charbonneau2001}. There {are suggestions} that continuum energy {relesase} models {can} translate to discrete self-organised critical models, 
such as those discussed in \cite{bak1996}. {2D numerical MHD models, which are restricted to direct forcing in the $x - y$ plane (e.g. \cite{dmitruk1998} 
and \cite{georgoulis1998}) lead to energy distributions
similar to that discussed by \cite{lu1991}}. {3D} MHD 
turbulent models of coronal heating have produced power law distributions of \lq events\rq , albeit over a limited range  \citep{rappazzo2007}, with the actual cause of the event 
sizes being subsumed into the general word \lq turbulence\rq .}

{So far, we have demonstrated that one magnetic thread can initiate energy release in a second, nearby, stable thread. It needs to be demonstrated that this merged flux tube can continue to disrupt a new (third) thread in its neighbourhood.} However,
it is difficult to run such a simulation involving many threads as sufficient grid resolution is required to ensure that numerical dissipation does not cause an 
individual thread to evolve in an unphysical manner.
Our investigations suggest that each thread must have at least 80 grid points across their diameters. Remembering that there may be potential fields between the threads 
and there must be a gap at the sides to reduce the influence
of the side boundary conditions, we can realistically simulate two (at high resolution) and possibly three magnetic threads (at lower resolution) with our present computing resources. Initial results for three threads
(at lower resolution) suggest that one unstable thread can destabilise only one other thread. However, there are now many different configurations to be investigated. In particular,
it is to be expected that the initial separation between the threads, relative sense of twist and initial field strength will determine whether these processes can release energy from many flux tubes. 
Such a release could describe a small flare. This will be discussed further in future publications.

In our simulations, each thread has the same basic form of magnetic field, with two parameters, namely the field strength on the thread axis and the twist parameter, $\lambda$. Both threads are twisted in the same sense. If
the twist in the second thread is reversed (basically reversing the direction of the axial current), for the case of touching threads, the stable thread does not destabilise. So the sense of twist could be important in whether an avalanche occurs or not. This is likely
to be because in this case the azimuthal field is in the same direction where the threads come into contact, mitigating against reconnection.

When the two threads are destabilised, the final relaxed state is effectively a single, large, weakly twisted loop. The length scales of this weakly twisted field are now larger than before, the diameter of the flux tube is about three and a half, whereas the original
diameters were each two. This is very similar to the inverse cascade of magnetic helicity reported by \cite{antiochos2013} and used by \cite{mackay2014}. We assume that if three threads can relax to form a single flux tube, then the length scales will increase again.

{What is the internal structure of coronal loops? The previous modelling of heating in kink unstable loops considers a single twisted cylindrical loop. Do they really consist of a single large cylindrical loop or are they made up of several (or many) smaller threads? 
The idea of multiple threads within
a coronal loop has been used by many 1D, single field line modellers \citep{cargill1994,klimchuk2008,bradshaw2011}. 
For these multi-thread loops, the coronal heating term is often specified as a function of position and time and is not, normally, determined in a self-consistent way, as in the 3D simulations. 
However, there are many benefits to the multi-thread models, in that they can reproduce the observed differential emission measures of coronal loops \citep{reale2014,warren2011,schmelz2013,cargill2014}.}

\section*{Acknowledgements}
We acknowledge the financial support of STFC through the Consolidated grant to the University of St Andrews. 
This work used the DIRAC 1, UKMHD Consortium machine at the University of St Andrews  and the DiRAC Data Centric system at Durham University, 
operated by the Institute for Computational Cosmology on behalf of the STFC DiRAC HPC Facility (www.dirac.ac.uk). This equipment was funded by BIS National
 E-infrastructure capital grant ST/K00042X/1, STFC capital grant ST/H008519/1, and STFC DiRAC Operations grant ST/K003267/1 and Durham University. DiRAC is part of the National E-Infrastructure.

\bibliographystyle{aa}
\bibliography{aa}
\end{document}